\documentclass[journal,twoside]{IEEEtran}

\usepackage[switch]{lineno}
\usepackage[utf8]{inputenc}
\usepackage{cite}
\usepackage{graphicx}
\usepackage{float,subcaption}
\usepackage{xargs}
\usepackage{mathrsfs}
\usepackage{amssymb,amsmath,amsthm,mathtools}
\usepackage{newtxmath}
\usepackage{siunitx}
\usepackage{textcomp}
\usepackage{xcolor,soul}
\definecolor{ao(english)}{rgb}{0.0, 0.5, 0.0}
\usepackage{hyperref}
\hypersetup{colorlinks, breaklinks, citecolor=blue, linkcolor=ao(english), urlcolor=blue}

\usepackage{cleveref}
\crefname{assumption}{Assumption}{Assumptions}

\usepackage{nicematrix}
\NiceMatrixOptions{code-for-first-row = \color{red},
	code-for-first-col = \color{blue},
	code-for-last-row = \color{green},
	code-for-last-col = \color{magenta}}

\usepackage{booktabs}
\usepackage{tabulary}
\usepackage{setspace}

\theoremstyle{plain}
\newtheorem{theorem}{Theorem}
\newtheorem{lemma}{Lemma}
\newtheorem{corollary}{Corollary}

\newtheorem{conjecture}{Conjecture}

\theoremstyle{remark}
\newtheorem{remark}{Remark}

\theoremstyle{definition}
\newtheorem{definition}{Definition}

\newtheorem*{problem*}{Problem}

\usepackage{cuted}

\usepackage{scalerel}
\usepackage{tikz}
\usetikzlibrary{svg.path}

\definecolor{orcidlogocol}{HTML}{A6CE39}
\tikzset{
	orcidlogo/.pic={
		\fill[orcidlogocol] svg{M256,128c0,70.7-57.3,128-128,128C57.3,256,0,198.7,0,128C0,57.3,57.3,0,128,0C198.7,0,256,57.3,256,128z};
		\fill[white] svg{M86.3,186.2H70.9V79.1h15.4v48.4V186.2z}
		svg{M108.9,79.1h41.6c39.6,0,57,28.3,57,53.6c0,27.5-21.5,53.6-56.8,53.6h-41.8V79.1z M124.3,172.4h24.5c34.9,0,42.9-26.5,42.9-39.7c0-21.5-13.7-39.7-43.7-39.7h-23.7V172.4z}
		svg{M88.7,56.8c0,5.5-4.5,10.1-10.1,10.1c-5.6,0-10.1-4.6-10.1-10.1c0-5.6,4.5-10.1,10.1-10.1C84.2,46.7,88.7,51.3,88.7,56.8z};
	}
}

\newcommand\orcidicon[1]{\href{https://orcid.org/#1}{\mbox{\scalerel*{
				\begin{tikzpicture}[yscale=-1,transform shape]
				\pic{orcidlogo};
				\end{tikzpicture}
			}{|}}}}

\usetikzlibrary{automata, shapes, arrows, calc, arrows.meta, fit, positioning}
\usepackage{pgfplots}
\usetikzlibrary{intersections}
\usetikzlibrary{graphs}

\begin{document}


\title{Consensus-driven Deviated Pursuit for Guaranteed Simultaneous Interception of Moving Targets}

\author{Abhinav~Sinha$^\dagger$\textsuperscript{\orcidicon{0000-0001-6419-2353}},~\IEEEmembership{Member,~IEEE},~Dwaipayan~Mukherjee\textsuperscript{\orcidicon{0000-0001-6993-9305}},~\IEEEmembership{Member,~IEEE}, and~Shashi~Ranjan~Kumar\textsuperscript{\orcidicon{0000-0001-6446-7281}},~\IEEEmembership{Senior Member,~IEEE}
\thanks{A. Sinha ($^\dagger$Corresponding author) is with the GALACxIS Lab, Department of Aerospace Engineering and Engineering Mechanics, University of Cincinnati, OH 45221, USA, e-mail: abhinav.sinha@uc.edu. D. Mukherjee is with Department of Electrical Engineering, Indian Institute of Technology Bombay, Powai- 400076, Mumbai, India, e-mail: dm@ee.iitb.ac.in. S.R. Kumar is with the Intelligent Systems \& Control Lab, Department of Aerospace Engineering, Indian Institute of Technology Bombay, Powai- 400076, Mumbai, India, e-mail: srk@aero.iitb.ac.in.}
}

%



\maketitle

\begin{abstract}
	This work proposes a cooperative strategy that employs deviated pursuit guidance to simultaneously intercept a moving (but not manoeuvring) target. As opposed to many existing cooperative guidance strategies which use estimates of time-to-go, based on proportional-navigation guidance, the proposed strategy uses an exact expression for time-to-go to ensure simultaneous interception. The guidance design considers nonlinear engagement kinematics, allowing the proposed strategy to remain effective over a large operating regime. Unlike existing strategies on simultaneous interception that achieve interception at the average value of their initial time-to-go estimates, this work provides flexibility in the choice of impact time. By judiciously choosing the edge weights of the communication network, a weighted consensus in time-to-go can be achieved. It has been shown that by allowing an edge weight to be negative, consensus in time-to-go can even be achieved for an impact time that lies outside the convex hull of the set of initial time-to-go values of the individual  interceptors. The bounds on such negative weights have been analysed for some special graphs, using Nyquist criterion. Simulations are provided to vindicate the efficacy of the proposed strategy.
\end{abstract}


%
\IEEEpeerreviewmaketitle

\section{Introduction}\label{sec:introduction}
\IEEEPARstart{E}{vidence} of cooperation is ubiquitous in natural entities and engineered systems. Trajectory shaping guidance strategies, e.g., \cite{10251969}, especially those that lead to simultaneous target interception, or \emph{salvo}, have gained prominence in recent years \cite{Jeon2010,9000526,8801941,Zhou2016,doi:10.2514/1.G005367,9447834}. This is mainly due to the advantages offered by introducing cooperation among interceptors, such as the ability to accomplish difficult missions easily, reduction in operational cost, increased operational reliability, etc. Interceptors' warhead effectiveness and their combat abilities can be enhanced if a swarm of interceptors are employed to simultaneously intercept a target as this may overwhelm the target's defence system. This, however, necessitates control of interceptors' impact time over a wide range of values.

Based on the target's mobility, that is whether the target is stationary or moving with a constant speed (henceforth \emph{moving}), the research in this direction can be broadly classified into two categories. Most of the research in the first category relies on using proportional-navigation (PN) guidance or its variants (e.g. see \cite{1597196,jeon2016impact}). 
It has been shown in \cite{Jeon2010,7126168,1597196,7376242,9000526,jeon2016impact} that the information of time-to-go is the key to designing impact time guidance strategies. \emph{Time-to-go}, a quantity associated with flight conditions in the future, is extremely difficult to obtain accurately when the interceptor is commanded by PN guidance. However, a reasonable estimate was derived in \cite{Jeon2010} to intercept a stationary target. A Lyapunov-based guidance was proposed in \cite{7126168} with an analysis of singularities in the guidance command. Authors in \cite{7376242,9000526} utilized sliding mode control to design guidance strategies, against a stationary target, that succeeded for engagements with very large heading angle errors. The aforementioned strategies, albeit good against stationary targets, were not originally developed to intercept moving targets. Further, most of them had some kind of singularity in the guidance command, which had to be carefully addressed. Some of these studies, \cite{Jeon2010,7376242,9000526}, considered the idea of predicted interception point (PIP), an approximation that considers a virtual stationary target, for extending the guidance strategy developed against stationary targets to intercept moving targets.  Vector guidance approach \cite{gutman2017impact} was also shown to be effective in intercepting moving targets, but  such a strategy demands radial acceleration, which is often difficult to realise in practice.

Apart from PN guidance, a variation of pursuit guidance, \emph{deviated-pursuit} \cite{shneydor1998missile}, was also noted in applications involving moving targets \cite{livermore2018deviated,segal2005derivation}. The work in \cite{kumar2019deviated} considered controlling time of interception using deviated pursuit. The motivation for using deviated pursuit guidance to intercept moving targets stems from the fact that it provides an exact expression for time-to-go against a target moving with constant velocity as it accommodates non-zero target velocity in its formulation. This not only leads to tractability of guidance design against such targets, but also ensures that the time-to-go, crucial in controlling impact time, is accurately known in closed form.

Most of the strategies discussed so far were developed for one-to-one engagements and may suffer a setback when pitted against powerful countermeasures employed by an adversary, such as close-in-weapon systems \cite{Jeon2010}. On the contrary, use of multiple interceptors were shown to be quite effective in simultaneous interception (e.g. see \cite{Jeon2010,9000526,8801941,Zhou2016,doi:10.2514/1.G005367} and references therein). A cooperative simultaneous interception can occur in two ways. Either the time of interception is commanded to each interceptor separately, and the interceptors do not communicate among themselves during the engagement, or the interceptors employ cooperative homing, in which the they communicate among themselves to synchronize their time of interception. Strategies, based on PN guidance, utilizing the former strategy have been reported in, e.g.,  \cite{Jeon2010,Zhou2016}. Note that if initial time-to-go is erroneous, simultaneous interception may fail in the absence of a corrective mechanism in this case, and such a system of interceptors rely on a centralised structure. Such a mechanism is also quite susceptible to homing anomalies and disturbances. The guidance strategies utilizing cooperative homing have the advantage that a simultaneous interception can still be achieved even if there is asynchronism in interceptors' launch timings or errors in initial time-to-go computations. This approach often uses the concept of consensus in multi-agent systems.

While most existing works (for instance \cite{7376242,9000526,LI2018243,Jeon2010,1597196}), have proposed cooperative guidance techniques to intercept a stationary target, intercepting a moving adversary is significantly more challenging. Moreover, the existing works did not consider the possibility of tailoring the consensus value of time-to-go, that is, intercepting the target simultaneously at an impact time other than the average of the interceptors' initial time-to-go values. In other words, these studies offered little flexibility in choosing the common impact time. This is a limitation since the desired common impact time need not always be the average of the interceptors' initial time-to-go. Further, from practical considerations, interceptors having larger time-to-go values at the beginning may demand considerably higher lateral acceleration (control effort) to achieve agreement with those interceptors whose initial time-to-go values are smaller. This may lead to acceleration saturation, or an actuator failure, eventually leading to mission failure. This motivates us to explore possibilities of having a simultaneous interception at an impact time, other than the average value of initial time-to-go.

Authors in \cite{8801941} proposed a cooperative guidance strategy in which the time of interception can be fixed a priori through a choice of gains. However, the scheme was restricted to communication over directed cycle graphs. In \cite{9000526,doi:10.2514/1.G005367}, a leader-follower cooperation was proposed in which the leader was explicitly aware of the desired time of interception, while followers used only relative time-to-go information to arrive simultaneously at the target. This implies over-dependence on the leader and leads to failure of mission in case the leader develops some snag. Further, the guidance schemes in \cite{8801941,9000526,doi:10.2514/1.G005367} were developed against stationary targets. To the best of the authors' knowledge, cooperative simultaneous interception schemes against moving targets, with the flexibility to tailor the time of interception, have not appeared in the literature so far. Some preliminary results in this direction were presented in our earlier work \cite{9147677}. The main contributions of this work, in view of the existing works, are summarized below.
\begin{itemize}
	\item A cooperative guidance strategy to intercept a moving target simultaneously using deviated pursuit guidance is proposed. Use of \emph{exact} time-to-go values eliminates errors that creep in due to approximations.
	\item Guidance design is carried out within a nonlinear framework, without any linearized approximations, thereby allowing the proposed scheme to remain effective for arbitrary initial engagement geometries, i.e., even for large heading angles of the interceptors. Furthermore, the proposed cooperative scheme does not have any singularities.
	\item We propose a special class of interaction topology, which we describe as a \emph{pseudo-undirected graph}, over which the interceptors communicate to achieve agreement in time-to-go. This offers a more general way of representing bidirectional information exchange among interceptors, with possibly different edge weights along the two directions of an undirected edge. 
	\item We show that a judicious selection of the edge weights in a {pseudo-undirected graph}, opens up the possibility of choosing the consensus value at will. Further, we illustrate that a single edge weight chosen to be negative (subject to a lower bound) enables consensus at a value outside the convex hull of the initial values of time-to-go. 
	\item Analysis of some special pseudo-undirected graphs has been carried out, which adds to the rich body of existing literature related to consensus in multi-agent systems. 
\end{itemize}

\section{Preliminaries and Problem Statement}\label{sec:background}
In this section, we present some preliminary concepts that aid in analysing the collective cooperative behaviour of the multi-interceptor swarm, followed by the problem formulation.
\subsection{Notation and Preliminaries}
We denote the set of real numbers by $\mathbb{R}$ and that of positive integers using $\mathbb{Z}_+$, while vectors are denoted using boldfaced letters in lower case. A column vector, $\vmathbb{1}_\mathrm{p}$, is the vector consisting of ones in $\mathbb{R}^\mathrm{p}$ whereas $\vmathbb{0}_\mathrm{p}$ is the vector containing zeros in $\mathbb{R}^\mathrm{p}$. For a matrix $\mathbf{M} = [m_\mathrm{ij}] \in \mathbb{R}^{\mathrm{p}\times \mathrm{p}}$, $\lambda_k(\mathbf{M})$ denotes its $k$\textsuperscript{th} eigenvalue where $k=\{1,\ldots,p\}$. A diagonal matrix, $\mathbf{M}$ is represented as $\mathbf{M}=\mathrm{diag}\left(\cdot\right)$, where the arguments are the ordered diagonal entries, while $\mathbf{I}_\mathrm{p}$ is the $p-$dimensional identity matrix. We denote the absolute value using $|\cdot|$ and 2-norm using  $\|\cdot\|$. We denote the convex hull of a set $\mathfrak{C} \subset \mathbb{R}$ by $\mathcal{C}o\{\mathfrak{C}\}$. For a matrix $\mathbf{M}$, the null space and the range space are $\mathfrak{N}(\mathbf{M})$ and $\mathcal{R}(\mathbf{M})$, respectively.  For a given set of vectors, $\mathfrak{s}\{\cdot\}$  represents the spanning set whereas $\mathfrak{b}\{\cdot\}$ represents the basis set. For a matrix $\mathbf{M}$, $\rho(\mathbf{M})$ denotes its rank. 

The cooperative behaviour of the multi-interceptor swarm is modelled via graphs with interceptors being represented as \emph{vertices} or \emph{nodes} and their interactions captured by \emph{edges}. An undirected simple graph \cite{mesbahi2010graph} is an ordered pair $\mathcal{G} = (\mathcal{V,E})$ consisting of a finite set of vertex (node) $\mathcal{V} = \{v_1, v_2, \ldots, v_n\}$ and an edge set $\mathcal{E} \subset \big\{\{v_i,v_j\} | v_i,v_j \in \mathcal{V} \;\text{and}\; i\neq j\big\}$. Cardinality of $\mathcal{V}$ is $n$ while that of $\mathcal{E}$ is usually considered to be $m$, unless otherwise specified. Later, when we introduce pseudo-undirected graphs, the number of edges, being always even, will be considered as 2$m$. Nodes $v_i, v_j \in \mathcal{V}$ are adjacent if $\{v_i, v_j\} \in \mathcal{E}$. For a node $v_i \in \mathcal{V}$, the degree of that node, $d(v_i)$, is the total number of edges entering or leaving $v_i$. The degree matrix, $\mathcal{D}(\mathcal{G})$, has  $d(v_i)$ on its main diagonal, that is, $[d_\mathrm{ij}]=d(v_i)$ if $i=j$, and $0$ otherwise. For an undirected graph, the elements of adjacency matrix, $\mathcal{A}(\mathcal{G})\in \mathbb{R}^{\mathrm{n}\times\mathrm{n}}$, are $[a_\mathrm{ij}]$ if $\{v_i,v_j\}\in \mathcal{E}$, and $0$ otherwise. The graph Laplacian, $\mathcal{L}(\mathcal{G})\in \mathbb{R}^{\mathrm{n}\times\mathrm{n}}$ is defined as $\mathcal{L}(\mathcal{G})=\mathcal{D}(\mathcal{G})-\mathcal{A}(\mathcal{G})$. An edge $\{v_i,v_j\}\in \mathcal{E}$ is said to be \emph{incident} on a node $v \in \mathcal{V}$ if $v = v_i$ or $v=v_j$. The incidence matrix, $E(\mathcal{G})\in \mathbb{R}^{\mathrm{n}\times\mathrm{m}}$ can be defined, after assigning random orientations to the undirected edges \cite{mesbahi2010graph}, as $[E(\mathcal{G})_\mathrm{ij}]=\pm 1$ if edge $e_j$ is incident on node $v_i$  (+1 indicates edge oriented away from node and vice versa), and zero otherwise. A \emph{path} may be described as a sequence of $k$ consecutively adjacent nodes, whose \emph{length} is given by $k-1$. The \emph{distance} between two vertices $v_i$ and $v_j$ in a graph $\mathcal{G}$ is the length of the shortest path between them, while the \emph{diameter}, $d$,  is the greatest distance between two nodes $v_i$ and $v_j$, $\forall i,j$.

For a directed graph (digraph), the edge set $\mathcal{E} \subset \mathcal{V} \times \mathcal{V}$, with each edge being represented as an ordered pair. An edge $e_\mathrm{ij}=(v_i,v_j)$ is said to exist if and only if the vertex $v_i$ receives the information from the vertex $v_j$, that is, $(v_i,v_j) \in \mathcal{E}$ represents a directed edge from $v_i$ to $v_j$. Thus, an edge represents a direction \emph{opposite} to the flow of information. Since the edges of a digraph are automatically oriented by the direction of the edge, the corresponding incidence matrix, $E(\mathcal{G})$, which is similar to that for an undirected graph, is described in terms of in- and out-incidence matrices ($E_\odot(\mathcal{G})$, with non-positive entries and $E_\otimes(\mathcal{G})$ with non-negative entries, which describe when an edge is incoming to or outgoing from a vertex, respectively) as $E(\mathcal{G})=E_\odot(\mathcal{G}) + E_\otimes(\mathcal{G})$, such that $E_\otimes(\mathcal{G})E_\otimes(\mathcal{G})^\top = \mathcal{D}(\mathcal{G})$, and $E_\otimes(\mathcal{G})E_\odot(\mathcal{G})^\top = -\mathcal{A}(\mathcal{G})$. Similarly, an out-Laplacian can be described as $\mathcal{L}_\otimes(\mathcal{G})=E_\otimes(\mathcal{G})E(\mathcal{G})^\top$ \cite{8444704}. Throughout this paper, we shall be dealing with such out-Laplacians for consensus over directed or pseudo-undirected graphs and hence omit the subscript $\otimes$ for Laplacians. For brevity, we shall also drop the arguments of all matrices when the underlying graph is clear from the context.  

Throughout this paper, we shall assume that the graphs are connected (undirected graphs) or have a directed spanning tree (digraphs) since in the absence of such structures, consensus cannot be achieved \cite{mesbahi2010graph}. For a digraph, the $(ij)$\textsuperscript{th} entry of $\mathcal{A}^k$  where $k\in\mathbb{Z}_+$ is nonzero if there exists a walk of length $k$ between vertices $v_i$ and $v_j$. The nature of $\mathcal{A}^k$ can be analysed using the notions of \emph{Perron}-\emph{Frobenius} theorem. Note that an eigenvalue of a matrix is called the dominant eigenvalue if it has the largest magnitude among all other eigenvalues.
\begin{definition}[\cite{noutsos2006perron}]
	A matrix $\mathbf{M} \in \mathbb{R}^{\mathrm{n}\times\mathrm{n}}$ is said to possess the \emph{Perron-Frobenius} property if its dominant eigenvalue, $\lambda_{1}$, is positive and the corresponding eigenvector is non-negative.
\end{definition}
\begin{definition}[\cite{noutsos2006perron}]
	A matrix $\mathbf{M} \in \mathbb{R}^{\mathrm{n}\times\mathrm{n}}$ is said to possess the \emph{strong Perron-Frobenius} property if its dominant eigenvalue, $\lambda_{1}$, is positive, with multiplicity one $(\lambda_{1}>\left|\lambda_{i}\right|, i= 2,3, \ldots, n)$, and the corresponding eigenvector is non-negative.
\end{definition}
\begin{definition}[\cite{noutsos2006perron}]
	A matrix $\mathbf{M} \in \mathbb{R}^{\mathrm{n}\times\mathrm{n}}$ is said to be \emph{eventually positive} (\emph{eventually non-negative}) if there exists a positive integer $z_0$ such that $\mathbf{M}^z$ is positive (non-negative) for all $z>z_0$.
\end{definition}
In certain situations, it may be convenient to associate a real number with each edge, i.e., we say that an edge has some weight. These weights signify the strength of information exchange between any two interceptors connected over an edge within the swarm. The  incidence matrix and the graph Laplacian of a weighted digraph are related as $\mathcal{L} = E_\otimes\mathcal{W}E^\top$, where $\mathcal{W}\in \mathbb{R}^{\mathrm{m}\times \mathrm{m}}$ is a diagonal matrix whose entries are the edge weights \cite{8444704}. The elements of a weighted adjacency matrix, $[a_\mathrm{ij}]$, now admit real entries and are no longer binary. Hence, for a weighted graph, $d(v_i) = \sum_{j}[a_\mathrm{ij}]$ are the diagonal entries of $\mathcal{D}(\mathcal{G})$.

In this work, we extend the notion of an undirected edge between a pair of nodes by viewing it as two oppositely directed edges between the same pair of nodes, with possibly \emph{different} weights along either direction. This is practically motivated as strength of the interaction may not be identical in both directions. This leads to a Laplacian that is no longer symmetric. Hence, for an undirected graph having $m$ edges, its corresponding pseudo-undirected graph has $2m$ edges. We now present the formal definition of a \emph{pseudo-undirected graph}.
\begin{definition}\label{def:pseudodirectedgraph}
	A pseudo-undirected graph, $\mathcal{G}$, is given by $(\mathcal{V,E,W})$, together with a map $\mathfrak{F}:\,\mathcal{E}\rightarrow\mathbb{R}^2$ such that for each pair $\{v_i,v_j\}\in \mathcal{E}$, we have $ \mathfrak{F}(e_\mathrm{ij},e_\mathrm{ji}) = [w_\mathrm{ij} ,w_\mathrm{ji}]^\top$ where $w_\mathrm{ij}$ and $w_\mathrm{ji}$ (entries of a diagonal matrix $\mathcal{W}\in\mathbb{R}^{2\mathrm{m}\times 2\mathrm{m}}$) are \emph{weights} corresponding to the directed edges $e_\mathrm{ij}$ and $e_\mathrm{ji}$, respectively.  Additionally, for such graphs, $w_\mathrm{ij} \neq w_\mathrm{ji}$ in general, and we need to ensure that $w_\mathrm{ij}w_\mathrm{ji}=0$ if and only if both $w_\mathrm{ij} = 0$ and $w_\mathrm{ji} = 0$.
\end{definition}
Therefore, for a pseudo-undirected graph, $\mathcal{G}$, we have $E_\otimes, E \in \mathbb{R}^{\mathrm{n}\times 2\mathrm{m}}$. Recall that the Laplacian for an undirected graph satisfies $\mathcal{L} = \mathcal{L}^\top$, and $\mathfrak{N}(\mathcal{L}) = \mathfrak{N}(\mathcal{L}^{\top})=\mathfrak{s}\{\vmathbb{1}_\mathrm{n}\}$. However, we only inherit the property $\mathfrak{N}(\mathcal{L}) =\mathfrak{s}\{\vmathbb{1}_\mathrm{n}\}$ for $\mathcal{G}$ while the remaining properties of an undirected graph need not hold.
\begin{definition}[\cite{9147677}]
	Consider an undirected graph obtained by replacing the non-zero edge weights of the pseudo-undirected graph by unity. The pseudo-undirected graph is said to be \emph{connected} if the corresponding equivalent undirected graph is connected.
\end{definition}
\begin{remark}
	The out degree of each node in the connected pseudo-undirected graph is at least one. Further, there exists a directed spanning subgraph, $\mathcal{G}_\tau \subseteq \mathcal{G}$ corresponding to at least one globally reachable node \cite{8444704}. For such graphs, there also exists a subgraph $\mathcal{G}_c$ such that $\mathcal{G}_\tau \cup \mathcal{G}_c = \mathcal{G}$. 
\end{remark}
\begin{remark}
	$\mathcal{G}_\tau$ contains $n-1$ directed edges, and the remaining $2m-n+1$ directed edges are contained in $\mathcal{G}_c$. Hence, the incidence matrix, $E(\mathcal{G})$, can be represented as
	\begin{equation}
		E(\mathcal{G})=\left[E(\mathcal{G}_\tau)\,E(\mathcal{G}_c)\right]=E(\mathcal{G}_\tau)[\mathbf{I}_{\mathrm{n}-1}\vdots-\mathbf{I}_{\mathrm{n}-1}\vdots\,\mathbf{T}_\tau\vdots-\mathbf{T}_\tau],
	\end{equation}
	where $E(\mathcal{G}_\tau)\in\mathbb{R}^{\mathrm{n}\times(\mathrm{n}-1)}$ and $E(\mathcal{G}_c)\in\mathbb{R}^{\mathrm{n}\times (2\mathrm{m}-\mathrm{n}+1)}$ capture the incidence relations in $\mathcal{G}_\tau$ and $\mathcal{G}_c$, respectively, and $\mathbf{T}_\tau$ is given by $\left(E^\top(\mathcal{G}_\tau)E(\mathcal{G}_\tau)\right)^{-1}E^\top(\mathcal{G}_\tau)E(\mathcal{G}_c)=[-\mathbf{I}_{\mathrm{n}-1}\vdots\,\mathbf{T}_\tau\vdots-\mathbf{T}_\tau]$. Conveniently, we may write $E(\mathcal{G}) = E(\mathcal{G}_\tau)\mathbf{R}$, where $\mathbf{R}=[\mathbf{I}_{\mathrm{n}-1}\,\vdots\,-\mathbf{I}_{\mathrm{n}-1}\,\vdots\, \mathbf{T}_\tau\, \vdots\, -\mathbf{T}_\tau]\in \mathbb{R}^{(\mathrm{n}-1)\times(2\mathrm{m}-\mathrm{n}+1)}$.
\end{remark}
From this point onward, we streamline our notation from $E(\mathcal{G}_\tau)$ to $E_\tau$ for brevity. Note that if the number of edges in a pseudo-undirected network equals twice the number of edges in $E_\tau$ (overall structure of corresponding undirected graph is that of a spanning tree), then $E=E_\tau[\mathbf{I}_{\mathrm{n}-1}\,-\mathbf{I}_{\mathrm{n}-1}]$.

The notion of incorporating a negative edge weight has shown promising applications, e.g., \cite{8638852,XIAO200465,XIA20112395}, but issues arising out of negative weights (e.g., robustness of consensus) need to be carefully addressed. The following lemmas, stated for directed graphs, aid in studying the bounds on negative edge weights for consensus protocols over pseudo-undirected graphs.
\begin{lemma}[\cite{8444704}]\label{lem:TF}
	Consider the uncertain edge agreement dynamics (uncertainty in the $i$-th edge) for single integrators in a rooted in-branching, $\dot{x}_\tau = -E_\tau^\top E_\otimes\left(\mathcal{W}+\mathbf{e}\Delta\mathbf{e}^\top\right)\mathbf{R}^\top x_\tau.$
	For a real, bounded and additive uncertainty, $\Delta<0$, the uncertain edge agreement protocol for single integrators is described by the transfer function
	\begin{equation}\label{eq:TF}
		M(s) = -\mathbf{e}^\top \mathbf{R}^\top\left[s\mathbf{I}+E_\tau^\top E_\otimes \mathcal{W} \mathbf{R}^\top\right]^{-1}E_\tau^\top E_\otimes \mathbf{e},
	\end{equation}
	where $-E_\tau^\top E_\otimes \mathcal{W} \mathbf{R}^\top$ is the system matrix, $-E_\tau^\top E_\otimes \mathbf{e}$ is the input matrix, $\mathbf{e}^\top \mathbf{R}^\top$ is the output matrix, $\mathbf{e}$ is the $i$\textsuperscript{th} standard basis in $\mathbb{R}^{2\mathrm{m}}$, used as an edge selection vector, and $\Delta\mathbf{e}^\top \mathbf{R}^\top x_\tau$ acts as input to the linear system with transfer function \eqref{eq:TF}.
\end{lemma}
\begin{lemma}[\cite{8444704}]\label{lem:GM}
	Consensus over pseudo-undirected graphs can be maintained if the uncertainty on a single edge weight is strictly less than the \emph{effective} gain margin of the transfer function, \eqref{eq:TF}, that is $|\Delta| < \frac{1}{|M(\jmath \omega_{pc})|}$,
	where $\omega_{pc}$ is the phase-crossover frequency corresponding to the smallest gain margin of $M(s)$.
\end{lemma}

\subsection{Problem Formulation}\label{sec:problemformulation}
\begin{figure}[h!]
	\centering
	\includegraphics[width=.5\linewidth]{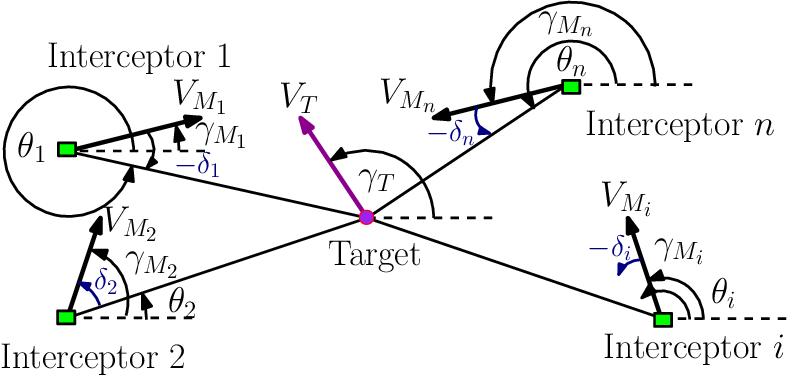}
	\caption{Planar engagement geometry for salvo interception.}
	\label{fig:eng_geo_multi}
\end{figure}

\Cref{fig:eng_geo_multi} depicts the many-to-one planar engagement geometry in which $n$ interceptors home onto a single moving target. For the $i$\textsuperscript{th} interceptor, $r_i$ and $\theta_i$, respectively, denote the relative range and the line-of-sight (LOS) angle, and $V_{\mathrm{M}_i}, V_\mathrm{T}$ are the speeds of the $i$\textsuperscript{th} interceptor, and the target, respectively. 
The deviation angle of the $i$\textsuperscript{th} interceptor is  $\delta_i$, while the flight path angles of the $i$\textsuperscript{th} interceptor and the target, are $\gamma_{\mathrm{M}_i}$ and $\gamma_\mathrm{T}$, respectively. The lateral acceleration of the $i$\textsuperscript{th} interceptor, which is its guidance command (control input), is represented by $a_{\mathrm{M}_i}$, with $V_{r_i}$ and $V_{\theta_i}$ being the respective relative velocity components along and perpendicular to the LOS. We assume that the vehicles are point mass vehicles, with constant speeds throughout the engagement, and that $V_{\mathrm{M}_i} > V_\mathrm{T}\;\forall\, i$. Further, we assume that there is no system lag due to autopilot dynamics. In the relative velocity space, the engagement kinematics of the $i$\textsuperscript{th} interceptor, with respect to a moving target, is governed by the set of following equations:
\begin{subequations}\label{eq:engkimenaticsdppmulti}
	\begin{align}
		V_{r_i}=\dot{r}_i =&~V_\mathrm{T} \cos (\gamma_\mathrm{T}-\theta_{i})-V_{\mathrm{M}_i}\cos\delta_i,\\
		V_{\theta_i}=r_i\dot{\theta}_i =&~V_\mathrm{T} \sin (\gamma_\mathrm{T}-\theta_{i}) -V_{\mathrm{M}_i}\sin\delta_i,\\
		\dot{\gamma}_{\mathrm{M}_i} =&~a_{\mathrm{M}_i}/V_{\mathrm{M}_i}.
	\end{align}
\end{subequations}
In \eqref{eq:engkimenaticsdppmulti}, the engagement kinematics is nonlinear, hence this model assumes no approximations due to linearization. To formulate the problem, we now present some necessary concepts related to impact time guidance and deviated pursuit.
	Note that the \emph{impact time}, $t_{\mathrm{im}}$, is the sum of expected outstanding flight time till interception, $t_{\mathrm{go}}$, and the time that has passed  after the interceptor's launch, $t_{\mathrm{el}}$, that is,  $t_{\mathrm{im}} =  t_{\mathrm{el}} + t_{\mathrm{go}}$.
\begin{lemma}[\cite{kumar2019deviated}]\label{tgolemma}
	The time-to-go for $i$\textsuperscript{th} interceptor in deviated pursuit, with constant angle of deviation, $\delta_i$, is given by
	\begin{align}
		t_{\mathrm{go}_i}= \dfrac{r_i}{\Upsilon_i}\left[V_{\mathrm{M}_i}\sec\delta_i+V_\mathrm{T}\sec\delta_i\cos\left(\gamma_\mathrm{T}-\theta_i+\delta_i\right)\right],\label{eq:tgodev}
	\end{align}
	where $\Upsilon_i=V_{\mathrm{M}_i}^2 - V_\mathrm{T}^2$, and the target is intercepted under deviated pursuit guidance strategy if and only if the time-to-go becomes zero.
\end{lemma} 
	The expression in \eqref{eq:tgodev} is exact, thus errors arising due to estimated values of time-to-go (common for PN based guidance strategies) are eliminated. 
We are now in a position to formally state the problem addressed in this work.
\begin{problem*}
	Design a cooperative guidance strategy to ensure simultaneous interception of a moving target over a wide range of common impact time value.
\end{problem*}
Since \Cref{tgolemma} presents an exact expression for time-to-go, it follows that a consensus in the same variable among the interceptors, in a cooperating group, suffices to ensure a simultaneous interception. In this regard, one may treat time-to-go as a state for the system described by \eqref{eq:engkimenaticsdppmulti}. It is also necessary to achieve a wide range of common impact time among interceptors that are possibly launched from different initial conditions.
\begin{remark}
	If the interceptors within the group communicate over an undirected graph whose edge-weights may be heterogeneous, the impact time, which is the consensus variable, will always be achieved at the average of the interceptors' initial time-to-go. On the other hand, if the interceptors are connected over a directed graph in which there are one or more globally reachable nodes or \emph{roots} (that is, nodes to which a path exists from every other node), then the consensus value is governed by the states of those interceptor(s) which constitute the roots in the graph. In either case, there will exist a subset of interceptors which do not influence the final consensus value and this may be a serious limitation if such interceptors are required to achieve a consensus value that is physically impossible. Such situations may arise when interceptors with larger initial time-to-go are required to achieve interception sooner than is physically possible and thus may demand higher acceleration (may even saturate), which could lead to failure of the salvo mission. 
\end{remark}

\section{Consensus over Pseudo-undirected Graphs}\label{sec:pug}
\begin{figure}[h!]
	\centering
	\includegraphics[width=\linewidth]{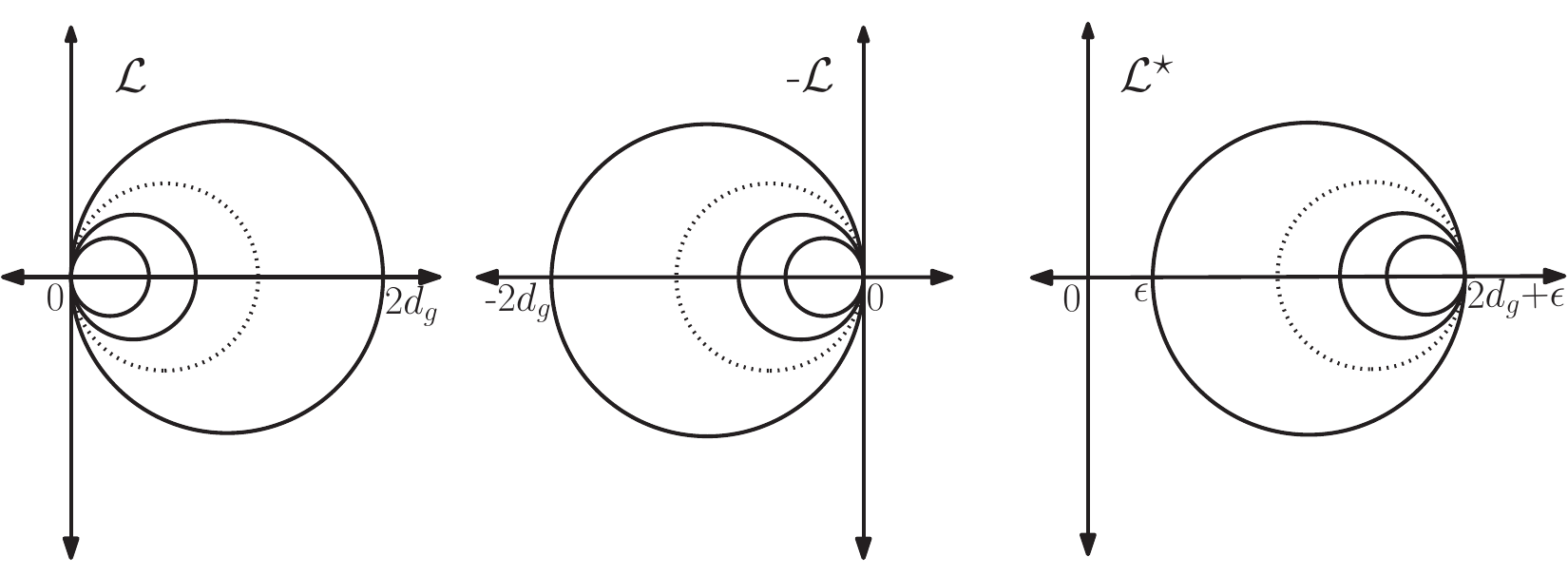}
	\caption{Ger\v{s}gorin regions for the eigenvalues of $\mathcal{L}$, $-\mathcal{L}$, $\mathcal{L}^\star$.}
	\label{fig:gdisk}
\end{figure}

Consider the matrix $\mathcal{L}^\star = \mathcal{A} - \mathcal{D} + (2d_g+\epsilon)\mathbf{I}_{\mathrm{n}}$, where $d_g = \max\limits_{i} d(v_i)$, and $\epsilon>0$. Typical Ger\v{s}gorin disks  for $\mathcal{L}$, $-\mathcal{L}$, and $\mathcal{L}^\star$ are shown in \Cref{fig:gdisk} for possibly non-identical, but positive edge weights.
\begin{lemma}\label{lem:Dstar}
	Consider the matrix $\mathcal{D}^\star = - \mathcal{D} + (2d_g+\epsilon)\mathbf{I}_{\mathrm{n}}$. The elements of $\mathcal{D}^\star$ are all non-negative.
\end{lemma}
\begin{proof}
	Since we know that $\max\limits_{i} |d_{ii}| = d_g$, therefore the smallest entry of $\mathcal{D}^\star$ is $2d_g+\epsilon-d_g = d_g+\epsilon>0$. Thus, $\mathcal{D}^\star$ is a diagonal matrix with all positive diagonal entries.
\end{proof}
\begin{lemma}\label{lem:Lstar}
	The matrix $\mathcal{L}^\star = \mathcal{A}+\mathcal{D}^\star$ is eventually positive.
\end{lemma}
\begin{proof}
	For any $k\in \mathbb{Z}_+$, $(\mathcal{A}+\mathcal{D}^\star)^k$ is given by $(\mathcal{A}+\mathcal{D}^\star)^k = \sum_{h=0}^{k}{k\choose h}\mathcal{A}^{k-h}\mathcal{D}^{\star^h}.$
	We know that the $(ij)$\textsuperscript{th} entry of $\mathcal{A}^{k-h}$ is positive if there is a walk of length $k-h$ from $v_i$ to $v_j$, and \Cref{lem:Dstar} ensures that $\mathcal{D}^\star$ has all positive entries on its diagonals. Since the diameter of $\mathcal{G}$ is $d$, a finite positive integer, we may conclude that for $k=d$, the matrix $(\mathcal{A}+\mathcal{D}^\star)$ has all positive entries. So, $\mathcal{L}^\star$ is an eventually positive matrix.
\end{proof}
\begin{lemma}\label{lem:ab}
	Two matrices $\mathbf{A}$ and $\mathbf{B}$ related as $\mathbf{A} + \Delta\mathbf{I}_\mathrm{n} = \mathbf{B}$, where $\Delta\in \mathbb{R}$, share the same eigenvectors up to a scaling.
\end{lemma}
\begin{proof}
	It readily follows that $\mathbf{A} + \Delta\mathbf{I}_\mathrm{n} = \mathbf{B}\Rightarrow \mathbf{A}\mathfrak{v} + \Delta\mathbf{I}_\mathrm{n}\mathfrak{v}= \mathbf{B}\mathfrak{v}\Rightarrow (\lambda_{\mathbf{A}} + \Delta)\mathfrak{v} = \mathbf{B}\mathfrak{v}$ where $\mathfrak{v}$ is an eigenvector of both $\mathbf{A}$ and $\mathbf{B}$ corresponding to eigenvalues $\lambda_{\mathbf{A}}$ and $\lambda_{\mathbf{A}} + \Delta$, respectively.
\end{proof}
\begin{lemma}\label{lem7}
	For a connected pseudo-undirected graph, $\mathcal{G}$, the eigenvector of $\mathcal{L}^\top$  for $\lambda(\mathcal{L}^{\top})=0$ is a positive vector.
\end{lemma}
\begin{proof}
	From \cite{noutsos2006perron} it follows that the properties of eventual positivity and strong Perron-Frobenius for $\mathcal{L}^{\star \top}$ and $\mathcal{L}^{\star}$ are equivalent. Hence, the eigenvectors of  $\mathcal{L}^{\star \top}$ and $\mathcal{L}^{\star}$ corresponding to the leading eigenvalue, $2d_g+\epsilon$, are positive. Using  \Cref{lem:ab}, one may write $\mathcal{L}^{\star \top} \mathfrak{v} = -\mathcal{L}^\top \mathfrak{v} + (2d_g+\epsilon)\mathfrak{v}$ for any eigenvector $\mathfrak{v}$ of either $-\mathcal{L}^\top$ or $\mathcal{L}^{\star \top}$. Corresponding to $\mathfrak{N}\{\mathcal{L}^\top\}$, it is immediate that $\mathcal{L}^{\star \top} \mathfrak{v} = (2d_g+\epsilon)\mathfrak{v}$. Hence, the vectors spanning $\mathfrak{N}\{\mathcal{L}\}$ (which is clearly $\vmathbb{1}_\mathrm{n}$) and $\mathfrak{N}\{\mathcal{L}^\top\}$ are both positive.
\end{proof}
For an undirected graph, $\mathcal{L} = \mathcal{L}^\top$, and $\mathfrak{N}(\mathcal{L}) = \mathfrak{N}(\mathcal{L}^{\top})=\mathfrak{s}\{\vmathbb{1}_\mathrm{n}\}$. 
However, since this is not necessarily true of pseudo-undirected graphs, our next objective is to evaluate the consensus value over such graphs, or at least provide a systematic method for obtaining the same. A key to this lies in obtaining the left null vector of the Laplacian matrix.
\begin{lemma}\label{lem:pw}
	Let us consider a non-trivial vector $\mathbf{p}\in\mathfrak{N}(\mathcal{L}^{\top})$. The following statements are equivalent. 1.) $\rho(E_\otimes) = n$, $\rho(\mathcal{W}) = 2m$, and $w_{ii}>0\,\forall\,i$, 2.) $(\mathbf{p}^\top E_\otimes \mathcal{W})^\top \in \mathfrak{N}(E)$.
	
\end{lemma}
\begin{proof}
	Recall that the weighted Laplacian can be expressed as $\mathcal{L} = E_\otimes \mathcal{W}E^\top$. Now, $\mathbf{p}\in \mathfrak{N}(\mathcal{L}^{\top})\implies\mathbf{p}\in  \mathfrak{N}(E\mathcal{W}E_\otimes^\top)\implies E\mathcal{W}E_\otimes^\top \mathbf{p} = \vmathbb{0}_\mathrm{n}$, from which it is immediate that $\mathbf{p}^\top E_\otimes \mathcal{W}E^\top = \vmathbb{0}^\top_\mathrm{n}$.
	A permutation of the columns of $E_\otimes$ reveals the identity matrix of dimension $n$ (since every node has at least an out-degree of unity). Thus, $E_\otimes\in\mathbb{R}^{\mathrm{n}\times 2\mathrm{m}}$ has full row rank of $n$, i.e., $\rho(E_\otimes) = n$. Now, $w_{ii}>0~\forall i$ implies that $\rho(\mathcal{W})=2m$. Hence, $\mathbf{p}^\top E_\otimes\neq \vmathbb{0}^\top_\mathrm{2m}$ and $ \mathbf{p}^\top E_\otimes \mathcal{W}\neq \vmathbb{0}^\top_\mathrm{2m}$, implying $(\mathbf{p}^\top E_\otimes \mathcal{W})^\top \in \mathfrak{N}(E)$. 
\end{proof}
\begin{lemma}\label{lem:p}
	For a vector $\mathbf{v} = \mathcal{W}E_\otimes^\top \mathbf{p}$, the following statements are equivalent.  1.) $\mathbf{v} \in \mathfrak{N}(E) \cap \mathcal{R}(\mathcal{W}E_\otimes^\top)$, 2.) $\mathbf{p} \in \mathfrak{N}(\mathcal{L}^\top)$.
\end{lemma}
\begin{proof}
	Since, $\mathbf{v} = \mathcal{W}E_\otimes^\top \mathbf{p}$, it is immediate that $\mathbf{v} \in\mathcal{R}(\mathcal{W}E_\otimes^\top)$. Pre-multiplying the vector $\mathbf{v}$ with the incidence matrix, we have $E\mathbf{v} = E\mathcal{W}E_\otimes^\top \mathbf{p}$, that is, $E\mathbf{v} = \mathcal{L}^\top\mathbf{p}$, which implies $\mathbf{v} \in \mathfrak{N}(E) \cap \mathcal{R}(\mathcal{W}E_\otimes^\top)\Longleftrightarrow \mathbf{p} \in \mathfrak{N}(\mathcal{L}^\top)$. 
\end{proof}
We know from \Cref{lem:p} that if we can find a $\mathbf{v}$ such that $\mathbf{v} \in \mathfrak{N}(E) \cap \mathcal{R}(\mathcal{W}E_\otimes^\top)$, we have the key to obtaining a vector $\mathbf{p} \in \mathfrak{N}(\mathcal{L}^\top)$. Since the existence of a non-trivial $\mathbf{p} \in \mathfrak{N}(\mathcal{L}^\top)$ is guaranteed due to the property of a graph Laplacian,  from \Cref{lem:pw,lem:p}, it follows that if we can constructively obtain the basis for $\mathfrak{N}(E) \cap \mathcal{R}(\mathcal{W}E_\otimes^\top)$, we can construct the vector  $ \mathbf{p}\in\mathfrak{N}(\mathcal{L}^\top)$, which in turn facilitates the evaluation of the consensus value.

Denote the basis sets $\bar{U} = \{u_1, u_2, \ldots, u_{2m-n+1}\}\in \mathfrak{b}\{\mathfrak{N}(E)\}$, and $\bar{V} = \{V_1, V_2, \ldots, V_n\} \in \mathfrak{b}\{\mathcal{R}(\mathcal{W}E_\otimes^\top)\}$. From \Cref{lem:p}, one can write $\mathbf{v} = \mathbf{U}x = \mathbf{V}y$, for some $x\in\mathbb{R}^{2\mathrm{m}-\mathrm{n}+1}$ and $y\in\mathbb{R}^\mathrm{n}$, where $\mathbf{U}=[u_1~\ldots ~u_{2m-n+1}]$ and $\mathbf{V}=[V_1~\ldots~ V_{n}]$.
\begin{remark}\label{rmk:Ux}
	Since, $\rho(\mathcal{W}E_\otimes^\top) = n$ (full column rank),  a possible choice of $\mathbf{V} = \left\{ [\mathcal{W}E_\otimes^\top]_1\;[\mathcal{W}E_\otimes^\top]_2\;\ldots\;[\mathcal{W}E_\otimes^\top]_n\right\}$, where the subscript denotes the corresponding column of $[\mathcal{W}E_\otimes^\top]$.
\end{remark}
\begin{remark}\label{rmk:Vy}
	$\rho(E) = n-1 \Rightarrow \mathfrak{N}(E)$ contains $2m-n+1$ vectors. One possible choice of $\mathbf{U}$ whose columns are a basis, $\mathfrak{b}\{\mathfrak{N}(E)\}$, is $$\begin{bmatrix}
		\mathbf{I}_{n-1} & \mathbf{T}_\tau & -\mathbf{T}_\tau \\
		\mathbf{I}_{n-1} & \mathbf{0} & \mathbf{0} \\
		\mathbf{0} & \mathbf{I}_{m-n+1} & \mathbf{0}\\
		\mathbf{0} & \mathbf{0} & \mathbf{I}_{m-n+1}\\
	\end{bmatrix},$$ which ensures full column rank of $\mathbf{U}$.
\end{remark}
From \Cref{rmk:Ux,rmk:Vy}, one has $x = (\mathbf{U}^\top \mathbf{U})^{-1}\mathbf{U}^\top \mathbf{V}y$, and $y = (\mathbf{V}^\top \mathbf{V})^{-1}\mathbf{V}^\top \mathbf{U}x$. Substituting for $y$ in the expression of $x$, we have $\mathbf{v} = \mathbf{U}x = \underbrace{\mathbf{U}(\mathbf{U}^\top \mathbf{U})^{-1}\mathbf{U}^\top}_{\mathbf{P}_\mathbf{U}}\underbrace{\mathbf{V}(\mathbf{V}^\top \mathbf{V})^{-1}\mathbf{V}^\top}_{\mathbf{P}_\mathbf{V}} \underbrace{\mathbf{U}x}_{\mathbf{v}}$. Thus, $\mathbf{v}=\mathbf{P}_\mathbf{U} \mathbf{P}_\mathbf{V} \mathbf{v} \in \mathfrak{N}(E) \cap \mathcal{R}(\mathcal{W}E_\otimes^\top)$.
Thus, the computation of $\mathbf{v}$ involves computing an eigenvector of the product of two projection matrices ${\mathbf{P}_\mathbf{U}}$ and ${\mathbf{P}_\mathbf{V}}$. Since, $\mathbf{v}=\mathbf{V}\mathbf{p}$, from \Cref{lem:p}, we may now obtain $\mathbf{p}=y=(\mathbf{V}^\top \mathbf{V})^{-1}\mathbf{V}^\top \mathbf{v}$.
\begin{remark}\label{rmk:consensusvalue}
	The consensus value is given by ${\sum_{i}p_{i}x_i(0)}/{\sum_{i}p_{i}}$ for the multi-agent systems communicating over a pseudo-undirected graph, whose dynamics is described by $\dot{\mathbf{x}}=-\mathcal{L}\mathbf{x}$, where $p_i$ is the $i$\textsuperscript{th} entry of $\mathbf{p}$, the vector spanning the null space of $\mathcal{L}$.
\end{remark}
\Cref{rmk:consensusvalue} shows that the consensus value, with heterogeneous edge weights, is at the weighted average of the initial states. As a result, the impact time is no longer required to be at the average of the initial time-to-go. By suitably chosen edge weights, the impact time, that is the consensus variable, can be made to converge to a desirable value, while respecting actuator constraints.
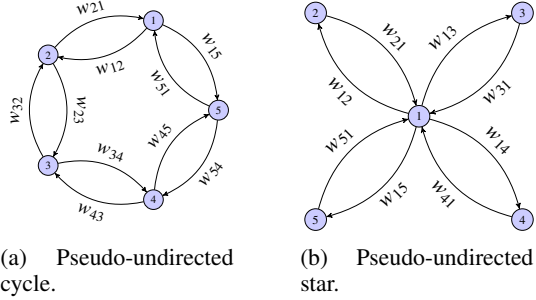
\begin{figure}[h!]
	\centering
	\begin{subfigure}[t]{.35\linewidth}
		\centering
		\resizebox{\textwidth}{!}{
			\begin{tikzpicture}
				\tikzstyle{vertex} = [circle,draw=black,fill=blue!20]
				\tikzstyle{edge} = [->,>=stealth',shorten >=0pt, bend left, thick, auto]
				\tikzstyle{dashedEdge} = [->,>=stealth',shorten >=0pt, bend left, thick, dotted, auto]
				\tikzstyle{writeBelow} = [sloped, anchor=center, below,scale=1.75]
				\tikzstyle{writeAbove} = [sloped, anchor=center, above,scale=1.75]
				\foreach \phi in {1,...,5}{
					\node[vertex] (v_\phi) at (360/5 * \phi:3cm) {$\phi$};
					
				}
				
				\draw[edge](v_1) edge node[writeBelow] {$w_{12}$} (v_2)
				(v_2) edge node[writeAbove] {$w_{21}$} (v_1) 
				(v_2) edge node[writeAbove] {$w_{23}$} (v_3)
				(v_3) edge node[writeAbove] {$w_{32}$} (v_2)
				(v_3) edge node[writeAbove] {$w_{34}$} (v_4)
				(v_4) edge node[writeBelow] {$w_{43}$} (v_3)		
				(v_4) edge node[writeAbove] {$w_{45}$} (v_5)
				(v_5) edge node[writeBelow] {$w_{54}$} (v_4)
				(v_5) edge node[writeBelow] {$w_{51}$} (v_1)
				(v_1) edge node[writeAbove] {$w_{15}$} (v_5)
				;		
			\end{tikzpicture}%
		}
		\caption{Pseudo-undirected cycle.}
		\label{fig:PUGcycle}
	\end{subfigure}\hfil
	\begin{subfigure}[t]{.35\linewidth}
		\centering
		\resizebox{\textwidth}{!}{
			\begin{tikzpicture}
				\tikzstyle{vertex} = [circle,draw=black,fill=blue!20]
				\tikzstyle{selectedVertex} = [circle,draw=black,fill=blue!20]
				\tikzstyle{edge} = [->,>=stealth',shorten >=0pt, bend left, thick, auto]
				\tikzstyle{writeBelow} = [sloped, anchor=center, below,scale=1.75]
				\tikzstyle{writeAbove} = [sloped, anchor=center, above,scale=1.75]
				
				\node[selectedVertex](1) at (0,0) {1};
				\node[vertex](2) at (-3,3) {2};
				\node[vertex](3) at (3,3) {3};
				\node[vertex](4) at (3,-3) {4};
				\node[vertex](5) at (-3,-3) {5};
				
				\draw[edge](1) edge node[writeBelow] {$w_{12}$} (2)
				(2) edge node[writeAbove] {$w_{21}$} (1) 
				(1) edge node[writeAbove] {$w_{13}$} (3)
				(3) edge node[writeBelow] {$w_{31}$} (1)
				(1) edge node[writeAbove] {$w_{14}$} (4)
				(4) edge node[writeBelow] {$w_{41}$} (1)
				(1) edge node[writeBelow] {$w_{15}$} (5)
				(5) edge node[writeAbove] {$w_{51}$} (1)
				;
			\end{tikzpicture}%
		}
		\caption{Pseudo-undirected star.}
		\label{fig:PUGstar}
	\end{subfigure}
	\caption{Special pseudo-undirected graphs.}
	\label{fig:PUG}
\end{figure}

In general, the entries of $\mathbf{p}$ depend upon the edge weights. When the edge weights are positive, the consensus value always lies within the convex hull of the initial states as established in \Cref{lem7}, that is, $\mathcal{C}o\{\mathfrak{C}\}$, where $\mathfrak{C}$ denotes the set containing initial time-to-go values of all interceptors. In practice, always having an impact time restricted to a given time interval has its limitations, as discussed earlier. In this work, we therefore explore the possibility of expanding the range of achievable impact time values beyond $\mathcal{C}o\{\mathfrak{C}\}$.
To do so, we next investigate the use of negative edge weight(s), subject to some lower bound (determined by the limiting case of failure to achieve consensus). 

\subsection{Pseudo-undirected Cycle}
A pseudo-undirected cycle with $5$ nodes ($n=5$) and $10$ edges ($m=5$) is shown in \Cref{fig:PUGcycle}. In general, the number of edges in a pseudo-undirected cycle is twice the number of nodes. The pseudo-undirected cycle has identical unweighted in-degree and out-degree equal to $2$, though with real valued weights, these are modified. The incidence matrix corresponding to a directed spanning subgraph, $E_\tau\in \mathbb{R}^{\mathrm{n}\times \mathrm{n}-1}$ can be given as
\begin{equation}\label{inci}
	E_\tau =\begin{pNiceMatrix}[nullify-dots]
		1 	    &  0      &  \Cdots &    	&  0 		\\
		-1 	    &  \Ddots &  \Ddots &  		&  \Vdots	\\
		0      & \Ddots  &         &  		& 			\\
		\Vdots &  \Ddots &         &  		&  0		\\
		&         &  		&  		&  1		\\
		0      & \Cdots  &  		&  0 	& -1
	\end{pNiceMatrix},
\end{equation}
and $\mathbf{T}_{\tau}=\vmathbb{1}_{\mathrm{n-1}}$. The weight matrix, $\mathcal{W}$, for such a graph is written as $\mathcal{W} = \mathrm{diag}(w_{1,2},\ldots, w_{n-1,n}, w_{2,1},\allowbreak \ldots,w_{n,n-1}, w_{n,1}, w_{1,n})$. The graph Laplacian, $\mathcal{L}$, corresponding to a pseudo-undirected cycle is given by
\begin{equation}\label{eq:Lcycle}
				\mathcal{L} = \begin{pmatrix}
						w_{1,2}+w_{1,n} & -w_{1,2} &  0&\cdots & -w_{1,n} \\
						-w_{2,1} & w_{2,1}+w_{2,3} & -w_{2,3}  & \cdots & 0  \\
						\vdots & \vdots & \vdots & \ddots &  \vdots   \\ 
						-w_{n,1} & 0 & 0 & \cdots & w_{n,1}+w_{n,n-1}
					\end{pmatrix}.
	\end{equation}
From \eqref{inci}, it is evident that $\mathfrak{N}(\mathcal{L})=\mathfrak{s}(\vmathbb{1}_\mathrm{n})$, but $\mathfrak{N}(\mathcal{L}^\top)\neq\mathfrak{s}(\vmathbb{1}_\mathrm{n})$ as $w_\mathrm{ij}\neq w_\mathrm{ji}$. Since the consensus over a pseudo-undirected graph occurs at a value given by ${\sum_{i}p_{i}x_i(0)}/{\sum_{i}p_{i}}$, we next endeavour to compute the elements of $\mathbf{p}\in\mathfrak{N}(\mathcal{L}^\top)$ when the interceptors are connected over a pseudo-undirected cycle.
\begin{lemma}\label{piCycle}
	For a pseudo-undirected cycle with $n\geq 4$, whose Laplacian is given by \eqref{inci}, the elements of $\mathbf{p}\in\mathfrak{N}(\mathcal{L}^\top)$ are given by
	\begin{equation}\label{eq:piCycle}
				\resizebox{\columnwidth}{!}{%
						$
			p_i = \dfrac{\displaystyle\prod_{\ell=i-1}^{i-(n-1)}w_{\ell,\ell-1}+\left(\sum_{b=1}^{n-1}\prod_{\ell=i}^{i-(b-1)}w_{\ell-1,\ell}\prod_{\ell=i-(b+1)}^{i-(n-1)}w_{\ell,\ell-1}\right)+\prod_{\ell=i-1}^{i-(n-1)}w_{\ell,\ell+1}}{\displaystyle\left(\prod_{\ell=1}^{n}w_{\ell,\ell+1}\right)\left(\prod_{\ell=1}^{n}w_{\ell,\ell-1}\right)},
						$%
					}
	\end{equation}
	where $i=1,2,\ldots,n$ and $\ell$ is computed modulo $n$ for $\ell>n$.
\end{lemma}
\begin{proof}
	A direct computation of the elements of $\mathbf{p}$, using \eqref{eq:piCycle}, and subsequent pre-multiplication of $\mathcal{L}$, as in \eqref{inci}, by this $\mathbf{p}$ so obtained completes the proof. The detailed calculations were omitted due to space constraints. 
\end{proof}	
However, as an illustration, we shall consider the pseudo-undirected cycle with $5$ nodes, as shown in \Cref{fig:PUGcycle}. Using \eqref{inci}, one may obtain $\mathcal{L}$ for the pseudo-undirected cycle and upon post-multiplying $\mathcal{L}$ 
with $\mathbf{p}=[p_1,p_2,p_3,p_4,p_5]$, and equating the product to $\vmathbb{0}_\mathrm{5}$, one may compute the entries of $\mathbf{p}$ as
\begin{align}
	p_1 =& \dfrac{1}{\left(w_{12}w_{23}w_{34}w_{45}w_{51}\right)w_{15}} +  \dfrac{1}{\left(w_{12}w_{23}w_{34}w_{45}\right)\left(w_{15}w_{54}\right)} \nonumber\\
	&+ \dfrac{1}{\left(w_{12}w_{23}w_{34}\right)\left(w_{15}w_{54}w_{43}\right)} + \dfrac{1}{\left(w_{12}w_{23}\right)\left(w_{15}w_{54}w_{43}w_{32}\right)} \nonumber\\
	&+ \dfrac{1}{w_{12}\left(w_{15}w_{54}w_{43}w_{32}w_{21}\right)},\nonumber \\
	p_2 =& \dfrac{1}{\left(w_{12}w_{23}w_{34}w_{45}w_{51}\right)w_{21}} + \dfrac{1}{\left(w_{23}w_{34}w_{45}w_{51}\right)\left(w_{15}w_{21}\right)}\nonumber\\
	& + \dfrac{1}{\left(w_{23}w_{34}w_{45}\right)\left(w_{15}w_{54}w_{21}\right)} + \dfrac{1}{\left(w_{23}w_{34}\right)\left(w_{15}w_{54}w_{43}w_{21}\right)} \nonumber\\
	&+ \dfrac{1}{w_{23}\left(w_{15}w_{54}w_{43}w_{32}w_{21}\right)} , \nonumber \\
	p_3 =&  \dfrac{1}{\left(w_{12}w_{23}w_{34}w_{45}w_{51}\right)w_{32}} + \dfrac{1}{\left(w_{12}w_{34}w_{45}w_{51}\right)\left(w_{32}w_{21}\right)}\nonumber\\
	& + \dfrac{1}{\left(w_{34}w_{45}w_{51}\right)\left(w_{15}w_{32}w_{21}\right)} + \dfrac{1}{\left(w_{34}w_{45}\right)\left(w_{15}w_{54}w_{32}w_{21}\right)} \nonumber\\
	&+ \dfrac{1}{w_{34}\left(w_{15}w_{54}w_{43}w_{32}w_{21}\right)} , \nonumber \\
	p_4 =& \dfrac{1}{\left(w_{12}w_{23}w_{34}w_{45}w_{51}\right)w_{43}} +  \dfrac{1}{\left(w_{12}w_{23}w_{45}w_{51}\right)\left(w_{43}w_{32}\right)}\nonumber\\
	& +  \dfrac{1}{\left(w_{12}w_{45}w_{51}\right)\left(w_{43}w_{32}w_{21}\right)}  + \dfrac{1}{\left(w_{45}w_{51}\right)\left(w_{15}w_{43}w_{32}w_{21}\right)} \nonumber\\
	&+ \dfrac{1}{w_{45}\left(w_{15}w_{54}w_{43}w_{32}w_{21}\right)},  \nonumber \\
	p_5 =&  \dfrac{1}{\left(w_{12}w_{23}w_{34}w_{45}w_{51}\right)w_{54}} + \dfrac{1}{\left(w_{12}w_{23}w_{34}w_{51}\right)\left(w_{54}w_{43}\right)} \nonumber\\
	&+
	\dfrac{1}{\left(w_{12}w_{23}w_{51}\right)\left(w_{54}w_{43}w_{32}\right)} +  \dfrac{1}{\left(w_{12}w_{51}\right)\left(w_{54}w_{43}w_{32}w_{21}\right)} \nonumber\\
	&+ \dfrac{1}{w_{51}\left(w_{15}w_{54}w_{43}w_{32}w_{21}\right)}, \label{eq:pcycle}
\end{align}
thereby revealing the pattern.

It is clear that the entries of $\mathbf{p}$ depend upon the edge weights, which in turn, govern the consensus value. Hence, altering an edge weight, to the point of making it negative will change the consensus value. For a weighted undirected network, the bounds on such negative edge weights were studied in \cite{7286773} and expressed in terms of \emph{graph resistances}, while the work in \cite{8444704} was focused on similar studies for weighted digraphs. Along similar lines, we analyse the bounds on negative edge weight values for networks represented by a pseudo-undirected graph. Without loss of generality, we shall assume that the edge $e_{12}$, with weight $w_{12}$, is perturbed. If $e_{12}$ is considered the first edge, then the edge selection vector becomes $\mathbf{e}=\left[1\,0\,0\,0\,0\,0\,0\,0\right]^\top$. Using \Cref{lem:TF}, the transfer function corresponding to the edge $e_{12}$ in the pseudo-undirected cycle, shown in \Cref{fig:PUGcycle}, is given as $M(s)=-N(s)/D(s)$, where $N(s)$ is a cubic monic polynomial and $D(s)$ (also monic) has degree four.

According to \Cref{lem:GM}, the maximum allowable perturbation $e_{12}$ can tolerate without violating consensus depends on the gain margin of the above transfer function, whose computation is our next objective. $M(s)$ can be compactly written as
\begin{align*}
	-M(s)=&\dfrac{N(s)}{D(s)} = \dfrac{s^3 + a_0 s^2 +a_1 s + a_2}{s^4 + b_0s^3 + b_1 s^2 + b_2 s +b_3}\\
	\implies -M(\jmath\omega)=& \dfrac{\left(a_2-a_0\omega^2\right) + j\left(a_1\omega-\omega^3\right)}{\left(\omega^4-b_1\omega^2+b_3\right) + j\left(b_2\omega - b_0\omega^3\right)},
\end{align*}
where $a_0,a_1,a_2,b_0,b_1,b_2,b_3$ are the coefficients of various powers of the complex variable $s$ as expressed above. Therefore, in order to find phase-crossover frequency, we compute $\angle -M(\jmath\omega)$ and equate it to $-\pi$, resulting in
\begin{equation*}
		\resizebox{\columnwidth}{!}{%
				$
		\tan^{-1} \dfrac{\left(a_1\omega-\omega^3\right)\left(\omega^4-b_1\omega^2+b_3\right)-\left(a_2-a_0\omega^2\right)\left(b_2\omega - b_0\omega^3\right)}{\left(a_2-a_0\omega^2\right)\left(\omega^4-b_1\omega^2+b_3\right)+\left(a_1\omega-\omega^3\right)\left(b_2\omega - b_0\omega^3\right)}=-\pi,
				$%
			}
\end{equation*}
from which we compute the phase-crossover frequency, $\omega_{pc}$, as $\omega_{pc}=0$ rad/s or the solution of 
\begin{equation*}
		\resizebox{\columnwidth}{!}{%
				$
		\omega_{pc}^6 + \left(a_0b_0-a_1-b_1\right)\omega_{pc}^4 + \left(b_3+a_1b_1-a_0b_2-a_2b_0\right)\omega_{pc}^2 + \left(a_2b_2-a_1b_3\right) = 0.
				$%
			}
\end{equation*}
\begin{remark}
	For a pseudo-undirected cycle with $5$ nodes, the computation of $\omega_{pc}$ involves solving a cubic polynomial in $\omega_{pc}^2$, which is still solvable in closed form. In general, however, as the number of nodes, $n$, increases, the degree of the polynomial also increases. Explicit closed-form solutions for polynomials of order beyond $4$ do not exist. Thus, the analytical computation of phase-crossover frequencies for a pseudo-undirected cycle is not feasible for large enough $n$, and we have to resort to graphical methods of analysis, such as Nyquist criterion. In general, one phase-crossover frequency is $\omega_{pc}=0$ rad/s and the rest are functions of edge weights, that is, $\omega_{pc}=f(w_\mathrm{ij})$.
\end{remark}
There exist multiple finite phase-crossover frequencies for pseudo-undirected graphs due to asymmetry in the edge weights of the forward and reverse edges between any two vertices, even though the graph topology is structurally symmetric. Here, we shall use Nyquist diagrams, a convenient tool to determine the stability margins of a linear system. As we shall show next, the maximum allowable perturbation on a single edge can be determined using Nyquist stability criterion from the transfer function corresponding to that edge. We next consider the special case where each edge weight is unity and compute the amount of perturbation any directed edge weight can tolerate.
\begin{conjecture}
	For the underlying undirected cycle with $n$ nodes, whose weights are homogeneous, that is, $w_{i,i+1}=w_{i+1,i}=1$ $\forall\,i$, $\omega_{pc}=0$ rad/s is the only phase-crossover frequency of the transfer function $M(s)$ corresponding to any edge, yielding a gain margin of $\frac{2}{1-\frac{1}{n}}$.
\end{conjecture}
\begin{proof}
	The underlying undirected cycle can be obtained by making $w_\mathrm{ij}=w_\mathrm{ji}=1~\forall i,j$, rendering the transfer function corresponding to all the edges same. As an example, consider the undirected cycle with $5$, $6$, and $7$ nodes obtained by substituting $w_\mathrm{ij}=w_\mathrm{ji}=1$ in the expression of $M(s)$. For an undirected cycle, whose weights are homogeneous, the expression of transfer function becomes
	{\small\begin{align*}
			-M(s)|_{n=5} =& \dfrac{s^3 + 7 s^2 + 15 s + 10}{s^4 + 10 s^3 + 35 s^2 + 50 s + 25} = \dfrac{s+2}{(s+3.618)(s+1.382)}, \\
			-M(s)|_{n=6} = & \dfrac{s^4+9s^3+28s^2+35s+15}{s^5+12s^4+54s^3+112s^2+105s+36} \\
			=& \dfrac{(s+3.618)(s+1.382)}{(s+4)(s+3)(s+1)}\\
			-M(s)|_{n=7} = & \dfrac{s^5+11s^4+45s^3+84s^2+70s+21}{s^6+14s^5+77s^4+210s^3+294s^2+196s+49} \\
			=& \dfrac{(s+3)(s+1)}{(s+3.8019)(s+2.445)(s+0.753)},
	\end{align*}}
	all of them have a single phase-crossover frequency at $0$ rad/s. Thus, gain margin is computed as $1/M(0)$. The gain margin, for any number of nodes, can be evaluated similarly by obtaining the corresponding transfer functions. Observe that the poles and zeroes of $M(s)$ are interlaced, and $1/M(0)$ has a sequence (from $n=3$ onward) $\dfrac{9}{3},\dfrac{16}{6},\dfrac{25}{10},\dfrac{36}{15},\dfrac{49}{21},\dfrac{64}{28},\dfrac{81}{36},\dfrac{100}{45},\ldots.$
	Using induction, the result can be generalized for $n$ nodes, resulting in a gain margin of $\frac{n^2}{\frac{n(n-1)}{2}}=\frac{2}{1-\frac{1}{n}}$.
\end{proof}
\begin{remark}
	Thus, for a pseudo-undirected cycle with identical edge weights, say unity, in all the $2m$ edges, $-\Delta<\frac{2}{1-\frac{1}{n}}$, that is, the minimum negative edge weight on any particular edge can be $w_\mathrm{ij}>1-\frac{2}{1-\frac{1}{n}} = -\frac{n+1}{n-1}$ must be maintained to avoid losing consensus. This means that for large number of agents, the robustness margin decreases and approaches two.
\end{remark}
\begin{figure}[h!]
	\centering
	\begin{subfigure}[t]{.475\columnwidth}
		\centering
		\includegraphics[width=\linewidth]{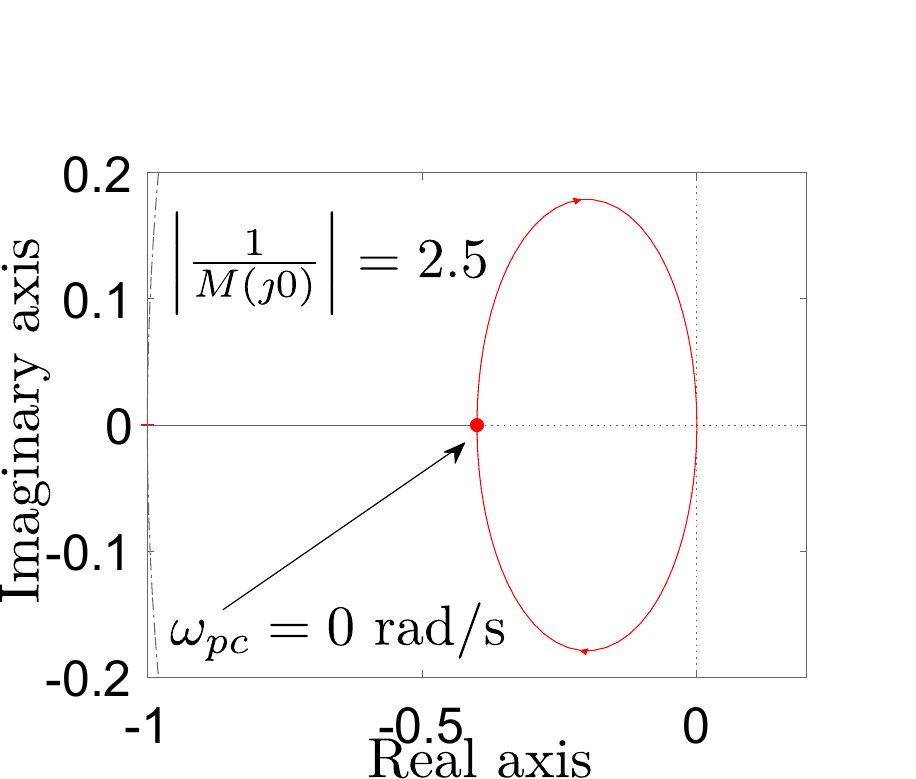}
		\caption{Homogeneous weights (all unity).}
	\end{subfigure}
	\begin{subfigure}[t]{.475\columnwidth}
		\centering
		\includegraphics[width=\linewidth]{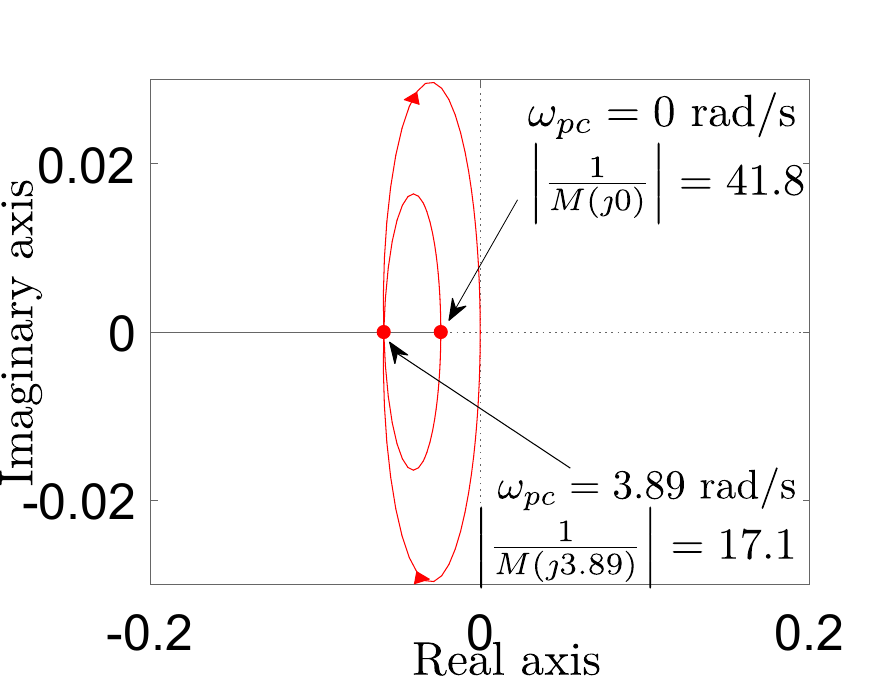}
		\caption{Heterogeneous weights.}
	\end{subfigure}
	\caption{Nyquist diagrams illustrating multiple $\omega_{pc}$.}
	\label{fig:Nyquistcycle}
\end{figure}
\Cref{fig:Nyquistcycle} shows the Nyquist diagrams for cases when the weights are all homogeneous and when $w_\mathrm{ij}\neq w_\mathrm{ji}$, as given in \Cref{tb:conensusheterogeneousnominal}.
\begin{table}[h!]
	\caption{Heterogeneous positive edge weights for pseudo-undirected cycle shown in  \Cref{fig:PUGcycle}.}\label{tb:conensusheterogeneousnominal}
	\centering
		\resizebox{\columnwidth}{!}{%
		\begin{tabular}{|c|c|c|c|c|c|c|c|c|c|}
			\hline
			$w_{12}$  & $w_{23}$ &$w_{34}$ & $w_{45}$ & $w_{51}$ & $w_{21}$  & $w_{32}$ &$w_{43}$ & $w_{54}$ & $w_{15}$ \\ \hline 
			$7$ & $3$ & $3$ & $8$ & $0.1$ & $0.3$ & $1.25$ & $5$ & $0.1$ & $5$ \\  \hline
		\end{tabular}%
			}
\end{table}
It can be inferred that the effective gain margin is determined by the leftmost phase-crossover and the corresponding frequency in case of heterogeneous weights, which in turn, dictates the maximum allowable perturbation on that particular edge, from its nominal positive value. For the weights given in \Cref{tb:conensusheterogeneousnominal}, the effective gain margin is $17.1$ and the weight assigned to $e_{12}$ is $w_{12}=7$, which essentially means that $w_{12}$ can be perturbed up to $-10.1$ without violating consensus.

\subsection{Pseudo-undirected Star}
A pseudo-undirected star with $5$ nodes is shown in \Cref{fig:PUGstar}. In this graph, the central node has an unweighted out-degree $n-1$, while rest of the nodes have out-degree $1$. For this topology, one may write
$E_\tau = [-\vmathbb{1}_{\mathrm{n}-1},\mathbf{I}_{\mathrm{n}-1}]^\top$ and $E=E_\tau\left[\mathbf{I}_{\mathrm{n}-1}\,-\mathbf{I}_{\mathrm{n}-1}\right]$. The edge weight matrix for a pseudo-undirected star is $\mathcal{W}=\mathrm{diag}(w_{1,2},\ldots, w_{1,n}, w_{n,1},\ldots,  w_{2,1})$, while its graph Laplacian is
\begin{equation}\label{eq:Lstar}
	\mathcal{L} =  \begin{pNiceMatrix}[nullify-dots]
		\sum_{\ell=2}^{n} w_{\ell,i} & -w_{i+1,i} &  \Cdots &  \Cdots & -w_{n,i} \\
		-w_{1,i} & w_{1,i} & 0 &  \Cdots & 0  \\
		\Vdots & 0 & \Ddots & \Ddots & \Vdots \\
		\Vdots & \Vdots & \Ddots & \Ddots & 0 \\ 
		-w_{1,n} & 0 & \Cdots & 0 & w_{1,n}
	\end{pNiceMatrix},
\end{equation}
where $i$ indexes the particular row. The eigenvector, $\mathbf{p}$, for a pseudo-undirected star can be computed as before, and is encapsulated in the result below.
\begin{lemma}
	The entries of $\mathbf{p}\in\mathfrak{N}(\mathcal{L}^\top)$ for a pseudo-undirected star, with $n\geq 2$ nodes, are given by 
	\begin{equation}\label{eq:pistar}
		p_1 = \displaystyle\prod_{\ell=2}^{n}w_{1,\ell},~p_i = p_1\dfrac{w_{i,1}}{w_{1,i}}\,\forall\,i\neq 1 .
	\end{equation}
\end{lemma}
\begin{proof}
	Similar to the proof of \Cref{piCycle}. 
\end{proof}
In a star network, every other node is connected to the central node only. As shown in case of pseudo undirected cycle, $M(s)$ may have multiple phase-crossover frequencies, which may not be possible to compute analytically. In order to generalize the essence of \Cref{lem:TF}, consider a perturbation on weight, $e_{12}$, for which we have $-M(s) = \mathbf{C}(s\mathbf{I}-\mathbf{A})^{-1}\mathbf{B} = \mathbf{C}\left(\dfrac{\mathrm{adjoint}(s\mathbf{I}-\mathbf{A})}{\mathrm{det}(s\mathbf{I}-\mathbf{A})}\right)\mathbf{B}$
as $(1,1)$\textsuperscript{th} entry of $(s\mathbf{I}-\mathbf{A})^{-1}$, with
{\small\begin{align*}
		\mathbf{A} =& \begin{bmatrix}
			-w_{12}-w_{21} & -w_{31} & -w_{41} &  \cdots &  \cdots & -w_{n,1}\\
			-w_{21} & -w_{13}-w_{31} & \vdots & \vdots & \vdots & \vdots \\
			\vdots & -w_{31} & \ddots & \vdots & \vdots &  \vdots \\
			\vdots & \vdots & \vdots & \ddots & \vdots & \vdots \\
			\vdots & \vdots & \vdots & \vdots & \ddots & -w_{n,1} \\
			-w_{21} & -w_{31} & -w_{41} & \cdots & \cdots & -w_{1,n}-w_{n,1} 
		\end{bmatrix},\displaybreak\\
		\mathbf{B} =&[-1,0,\ldots,0]^\top,~\mathbf{C} =[1,0,\ldots,0].
\end{align*}} 
Thus, for $n$ number of nodes,
\begin{equation}\label{eq:TFstar}
		\resizebox{\columnwidth}{!}{%
				$
		M(s) = -\dfrac{\det(s\mathbf{I}-\mathbf{A})_{(1,1)}}{(s+w_{12}+w_{21})\det(s\mathbf{I}-\mathbf{A})_{(1,1)}+w_{21}\sum_{\ell=2}^{n-1}(-1)^{\ell+1}\det(s\mathbf{I}-\mathbf{A})_{(\ell,1)}},
				$%
			}
\end{equation}
\begin{corollary} 
	For the underlying undirected star with $n$ nodes, whose weights are homogeneous, that is, $w_\mathrm{i1}=w_\mathrm{1i}=1$, $\omega_{pc}=0$ rad/s is the only phase-crossover frequency of the transfer function $M(s)$ corresponding to perturbation on edge, $e_{12}$, with a gain margin of $\dfrac{1}{1-\frac{1}{n}}$, while for a perturbation on edge $e_{21}$, this bound is $n$.
\end{corollary}
\begin{proof}
	For an equivalent undirected star with $n$ nodes, letting $w_\mathrm{ij}=w_\mathrm{ji}=1$, we have $\mathbf{A} = \begin{bmatrix*}[r]
		-2 & -1 & \cdots &  -1\\
		-1 & \ddots & \ddots & \vdots \\
		\vdots & \ddots & \ddots & -1  \\
		-1 & \cdots & -1 & -2
	\end{bmatrix*}$, $(s\mathbf{I}-\mathbf{A})^{-1} = \dfrac{1}{(s+1)(s+n)}\begin{bmatrix}
	s+n-1 & -1 & \cdots &  -1\\
	-1 & \ddots & \ddots & \vdots \\
	\vdots & \ddots & \ddots & -1  \\
	-1 & \cdots & -1 & s+n-1
\end{bmatrix}$,
	and it follows from \eqref{eq:TFstar} that $M(s) = -\dfrac{s+(n-1)}{(s+1)(s+n)}$,
	which has a single phase-crossover frequency, $\omega_{pc}=0$. Hence, the gain margin is computed as $\left\vert\frac{1}{M(0)}\right\vert=\frac{n}{n-1} = \frac{1}{1-\frac{1}{n}}$. Similarly, if $w_{21}$ is perturbed, then the system matrix, $\mathbf{A}$, remains unchanged. Thus, for identical unity edge weights, we have the transfer function, corresponding to perturbation on $w_{21}$, as
	\begin{equation}
		M(s) = \mathbf{C}(s\mathbf{I}-\mathbf{A})^{-1}\mathbf{B} = -\dfrac{1}{s+n}
	\end{equation}
	with $\mathbf{B} = \mathbf{1}_{n-1}$ and $\mathbf{C} =[-1,0,\ldots,0]$. Therefore, $M(s)$ so obtained has a single phase-crossover frequency, $\omega_{pc}=0$. Hence, the gain margin is computed as $\left\vert\dfrac{1}{M(0)}\right\vert=n$.
\end{proof}

In general, when $w_\mathrm{ij}\neq w_\mathrm{ji}$, and $w_{21}$ is perturbed, we have
\begin{equation}
		\resizebox{\columnwidth}{!}{%
				$
		M(s) = \dfrac{-\displaystyle\sum_{\ell=1}^{n-1}(-1)^{\ell+1}\det(s\mathbf{I}-\mathbf{A})_{(\ell,1)}}{(s+w_{12}+w_{21})\det(s\mathbf{I}-\mathbf{A})_{(1,1)}+w_{21}\displaystyle\sum_{\ell=2}^{n-1}(-1)^{\ell+1}\det(s\mathbf{I}-\mathbf{A})_{(\ell,1)}}.
				$%
			}
\end{equation}

\section{Design of the Guidance Strategy}\label{sec:guidance}
In this section, we shall derive the guidance strategy for a multi-interceptor group to simultaneously intercept a moving target, using the above results on weighted consensus.
\begin{lemma}\label{lem4}
	The dynamics of time-to-go for $i$\textsuperscript{th} interceptor guided by deviated pursuit guidance strategy can be written as
	\begin{align}
		\dot{t}_{\mathrm{go}_i}=&~-1+\dfrac{V_{\theta_i}^2\sec^2\delta_i}{\Upsilon_i}- \dfrac{r_iV_{\theta_i}\sec^2\delta_i}{V_{\mathrm{M}_i}\Upsilon_i^2} a_{\mathrm{M}_i}.\label{tgopurdyn}
	\end{align}
\end{lemma}
\begin{proof}
	On differentiating $t_{\mathrm{go}_i}$, as in \eqref{eq:tgodev}, with respect to time, we get
		\begin{align}
		\dot{t}_{\mathrm{go}_i}=&~ \dfrac{1}{\Upsilon_i}\left[V_{r_i}(V_{r_i}+2V_{\mathrm{M}_i}\cos\delta_i - V_{\theta_i}\tan\delta_i)+ r_i(\dot V_{r_i} \right.\nonumber\\
		&\left.-2V_{\mathrm{M}_i}\sin\delta_i\dot \delta_i-\dot V_{\theta_i}\tan\delta_i-V_{\theta_i}\sec^2\delta_i\dot\delta_i)\right].\label{tgodot}
	\end{align}
	Further differentiating \eqref{eq:engkimenaticsdppmulti} with respect to time, one obtains 
		$\dot V_{r_i} =~\frac{V_{\theta_i}^2}{r_i}+a_{\mathrm{M}_i}\sin\delta_i,~\dot V_{\theta_i} =~ -\frac{V_{r_i}V_{\theta_i}}{r_i}-a_{\mathrm{M}_i}\cos\delta_i,$
	which upon substitution in \eqref{tgodot}, along with the use of $\dot \delta_i=(a_{\mathrm{M}_i}/V_{\mathrm{M}_i})-\dot\theta_i$, results in \eqref{tgopurdyn}. This completes the proof.
\end{proof}
\begin{theorem}
	For the dynamics of $t_{\mathrm{go}_i}$ in \Cref{lem4}, the cooperative guidance strategy for $i$\textsuperscript{th} interceptor that ensures simultaneous interception of a moving target for arbitrary initial engagement geometry, is
	\begin{equation}\label{eq:aMform}
		a_{\mathrm{M}_i} =V_{\mathrm{M}_i}\dot{\theta}_i - \dfrac{V_{\mathrm{M}_i}\Upsilon_i \cos^2 \delta_i}{r_i V_{\theta_i}}\sum_{j}[l_\mathrm{ij}]t_{\mathrm{go}_j} ,
	\end{equation}
	where $[l_\mathrm{ij}]$ are the entries of the Laplacian matrix, $\mathcal{L}$.
\end{theorem}
\begin{proof}
	Let the error in interception time for $i$\textsuperscript{th} interceptor be $e_i(t) = t_{\mathrm{go}_i} - t_\mathrm{go}^\mathrm{d}$, where $t_\mathrm{go}^\mathrm{d}$ is some desired time-to-go that is to be cooperatively achieved. For a constant desired interception time, decided implicitly by a choice of edge weights, $\dot{t}_\mathrm{go}^\mathrm{d} = -1$. Thus,	
	\begin{align}
		\dot{e}_i(t) &= \dot{t}_{\mathrm{go}_i} + 1	\implies \dot{e}_i(t) = {F}_i + {B}_i a_{\mathrm{M}_i} + 1,\label{eq:ei}
	\end{align}
	where $F_i = -1+\frac{V_{\theta_i}^2\sec^2\delta_i}{\Upsilon_i^2}$ and $B_i = \frac{r_iV_{\theta_i}\sec^2\delta_i}{V_{\mathrm{M}_i}\Upsilon_i^2}$. If the cooperative guidance command for $i$\textsuperscript{th} interceptor is chosen as 
	\begin{align}\label{ami}
		a_{\mathrm{M}_i} &= -\dfrac{{F}_i + 1 - u_i}{{B}_i},
	\end{align}
	then the error dynamics, \eqref{eq:ei}, reduces to $\dot{e}_i(t) = u_i$, where $u_i$ is an auxiliary control input. With this single integrator dynamics, previously derived results on consensus can be readily applied.
	
	With $u_i = -\sum_{j}[l_\mathrm{ij}]e_j$, the error dynamics, \eqref{eq:ei}, becomes $\dot{e}_i = -\sum_{j}[l_\mathrm{ij}]e_j$, resulting in asymptotic consensus
	in ${e}_i$, and therefore, in $t_{\mathrm{go}_i}$. Substituting the expressions of ${F}_i$ and ${B}_i$ in \eqref{ami} results in the cooperative guidance command, \eqref{eq:aMform}. This concludes the proof.
\end{proof}
\begin{remark}
	Once the interceptors are in agreement with respect to their time-to-go, the term responsible for maintaining cooperation, $-\frac{V_{\mathrm{M}_i}\Upsilon_i \cos^2 \delta_i}{r_i V_{\theta_i}}\sum_{j}[l_\mathrm{ij}]e_j$, tends to zero. The cooperative guidance command, \eqref{eq:aMform}, then, becomes equivalent to $V_{\mathrm{M}_i}\dot{\theta}_i$ only, which is the term responsible for maintaining the interceptors on the pursuit course while their  deviation angles remain fixed.
\end{remark}
\begin{remark}
	Rather than leading to an average consensus, the guidance command \eqref{eq:aMform} leads to a weighted average consensus in time-to-go values (see \Cref{rmk:consensusvalue}) where the choice of edge weights dictate the consensus value. For positive weights, the consensus value is within the convex hull of initial time-to-go values, that is, within $\mathcal{C}o\{\mathfrak{C}\}$ where $\mathfrak{C}$ is the set containing the interceptors' initial time-to-go values. On the other hand, if one of the edge weights is negative but within the acceptable bound (see \Cref{lem:GM}), the time of interception may also lie outside $\mathcal{C}o\{\mathfrak{C}\}$. However, it is important to note that due to physical constraints, the minimum achievable impact time for an interceptor $i$ is $r_i/V_{\mathrm{M}_i}$.
\end{remark}
Having an impact time, that is the consensus value, outside $\mathcal{C}o\{\mathfrak{C}\}$ can enhance the applicability of the proposed cooperative guidance strategy as a wide range of common impact time values become feasible for the group of interceptors. This is shown via simulations in the next section.

\section{Simulations}\label{sec:simulations}
In order to show the efficacy and wide applicability of the proposed guidance scheme, we consider five interceptors communicating over pseudo-undirected versions of the special graphs discussed in the previous sections. The speed of each interceptor is $500$ m/s while that of the target is $400$ m/s. Initially, the interceptor and the target are $10$ km radially apart. We have assumed that the interceptors, owing to physical constraints on the actuators, can draw lateral accelerations no more than $40$ g, where g is the acceleration due to gravity. In the subsequent figures, the square markers in the trajectory plots correspond to the interceptors' initial positions, while pink circle markers appear on the target after every $10$ s of engagement. In each figure, $I_i$ corresponds to the $i$\textsuperscript{th} interceptor.

First of all, the proposed cooperative guidance strategy is demonstrated on a pseudo-undirected cycle graph, shown in \Cref{fig:PUGcycle}, with heterogeneous positive weights. From \Cref{rmk:consensusvalue}, it has been established that the impact time, which is the consensus value, lies at the weighted average of the interceptors' initial time-to-go (states), when the edge weights are heterogeneous. For the case of a pseudo-undirected cycle, \Cref{tb:conensusheterogeneousnominal} presents the edge weights, while the interceptors' initial engagement geometry is given in \Cref{tb:initCycle}.
\begin{table}[ht!]
	\caption{Parameters of simulation for pseudo-undirected cycle graph.}\label{tb:initCycle}
	\centering
	\begin{tabular}{|c|c|c|c|c|c|}
		\hline
		\phantom{-}& $I_1$  &$I_2$ & $I_3$ & $I_4$ & $I_5$\\ \hline 
		$\theta_i$ & $0^\circ$ & $-5^\circ$& $-10^\circ$& $-15^\circ$ & $-20^\circ$\\  \hline
		$\gamma_{\mathrm{M}_i}$ & $0^\circ$ & $10^\circ$ & $20^\circ$ & $-10^\circ$ & $-20^\circ$\\  \hline
		$t_{\mathrm{go}_i}(0)$ & $33.33$ s & $22.26$ s & $15.92$ s & $21.59$ s & $21.50$ s \\ \hline
	\end{tabular} 
\end{table}

\begin{figure*}[h!]
	\centering
	\begin{subfigure}[t]{.33\textwidth}
		\centering
		\includegraphics[width=\textwidth]{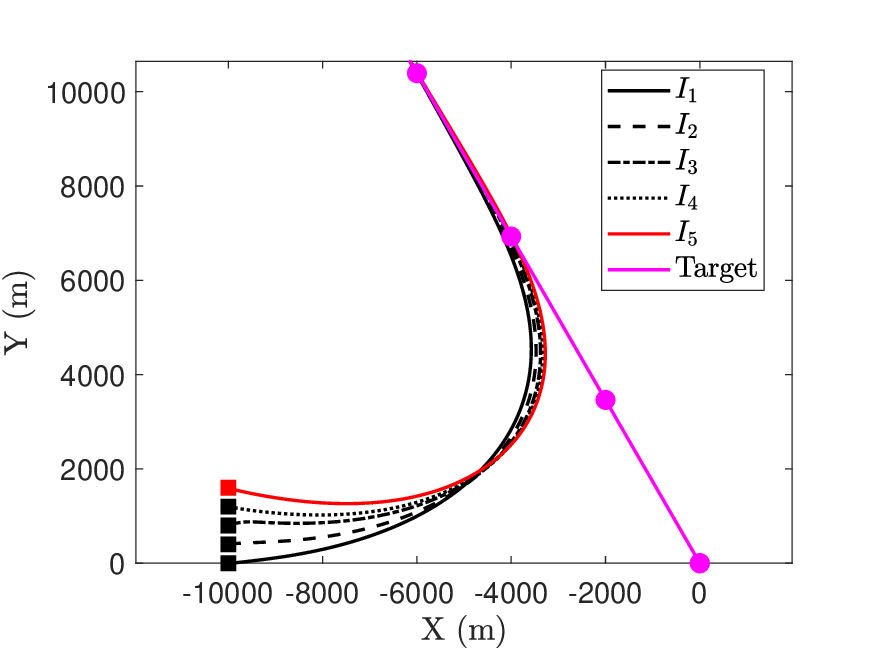}
		\caption{Trajectories.}
		\label{fig:cycleHePosTraj}
	\end{subfigure}
	\begin{subfigure}[t]{.32\textwidth}
		\centering
		\includegraphics[width=\textwidth]{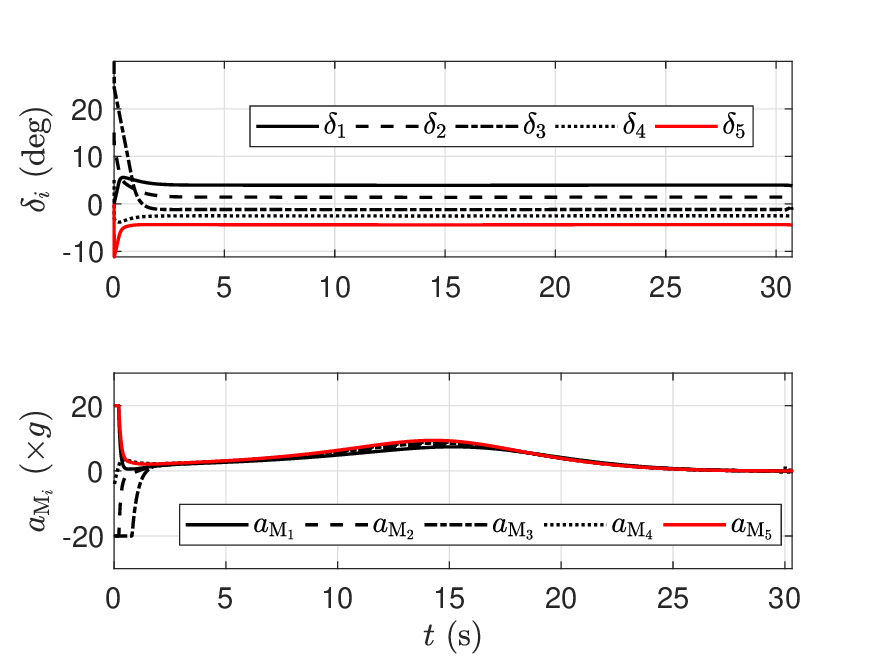}
		\caption{Deviation angles and lateral accelerations.}
		\label{fig:cycleHePosAm}
	\end{subfigure}
	\begin{subfigure}[t]{.33\textwidth}
		\centering
		\includegraphics[width=\textwidth]{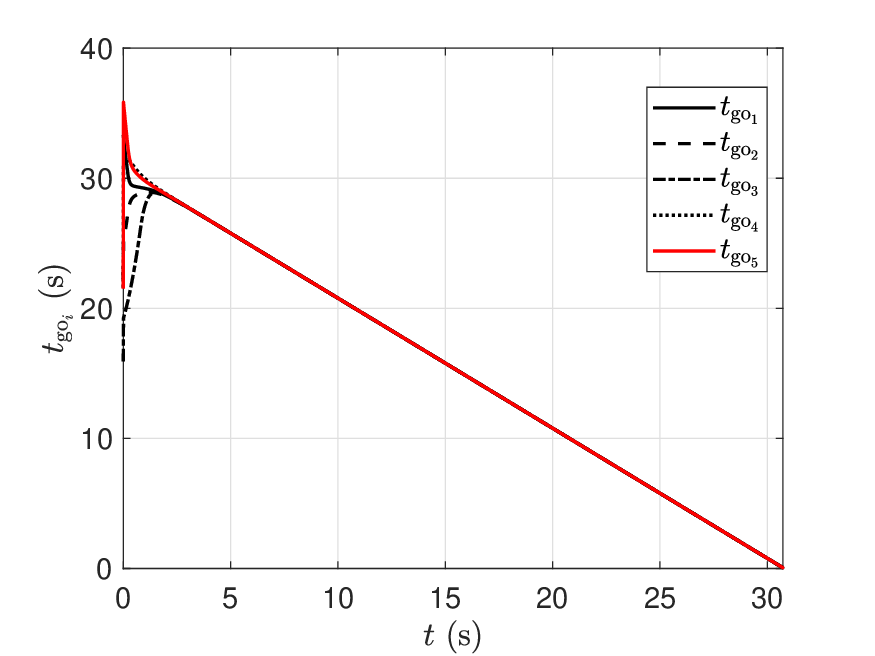}
		\caption{Time-to-go.}
		\label{fig:cycleHePosTgo}
	\end{subfigure}
	\caption{Weighted consensus in time-to-go (pseudo-undirected cycle).}
	\label{fig:cycleHePos}
\end{figure*}

\Cref{fig:cycleHePos} depicts cooperative simultaneous interception of a moving target at an impact time of $30.74$ s. The target maintains a constant heading of $120^\circ$. The trajectories of the participants are shown in \Cref{fig:cycleHePosTraj}, from which it can be observed that interceptors start close to each other with different launch headings, and intercept the moving target simultaneously. \Cref{fig:cycleHePosAm} shows the variation of the deviation angles and the lateral acceleration profiles of interceptors, while the time-to-go of each interceptor is presented in \Cref{fig:cycleHePosTgo}. It is inferred from \Cref{fig:cycleHePosAm,fig:cycleHePosTgo} that as soon as agreement in time-to-go is established, around $3$ s, the deviation angles becomes constant and the lateral accelerations tend towards zero. 
\begin{figure*}[h!]
	\centering
	\begin{subfigure}[t]{.33\textwidth}
		\centering
		\includegraphics[width=\textwidth]{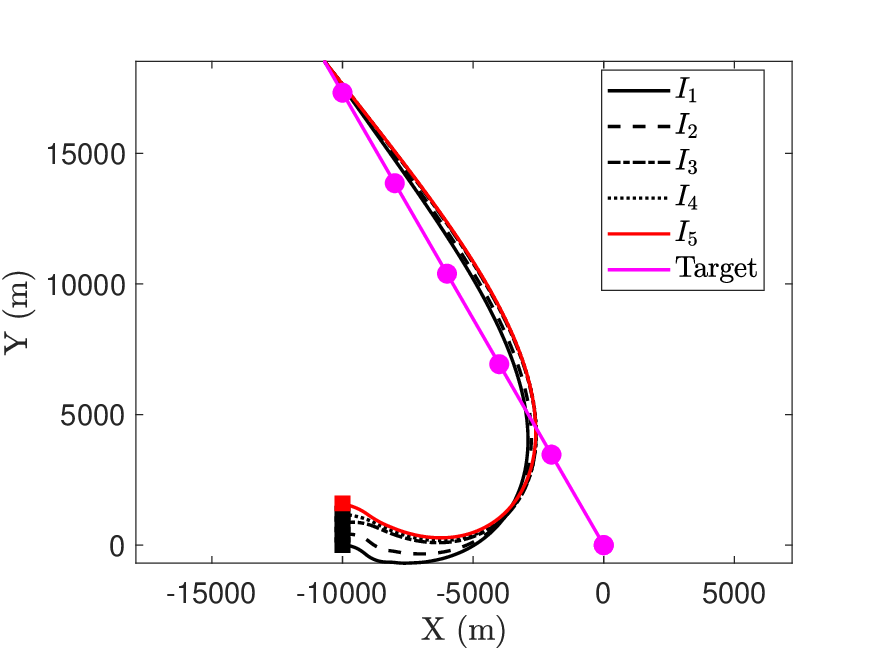}
		\caption{Trajectories.}
	\end{subfigure}
	\begin{subfigure}[t]{.32\textwidth}
		\centering
		\includegraphics[width=\textwidth]{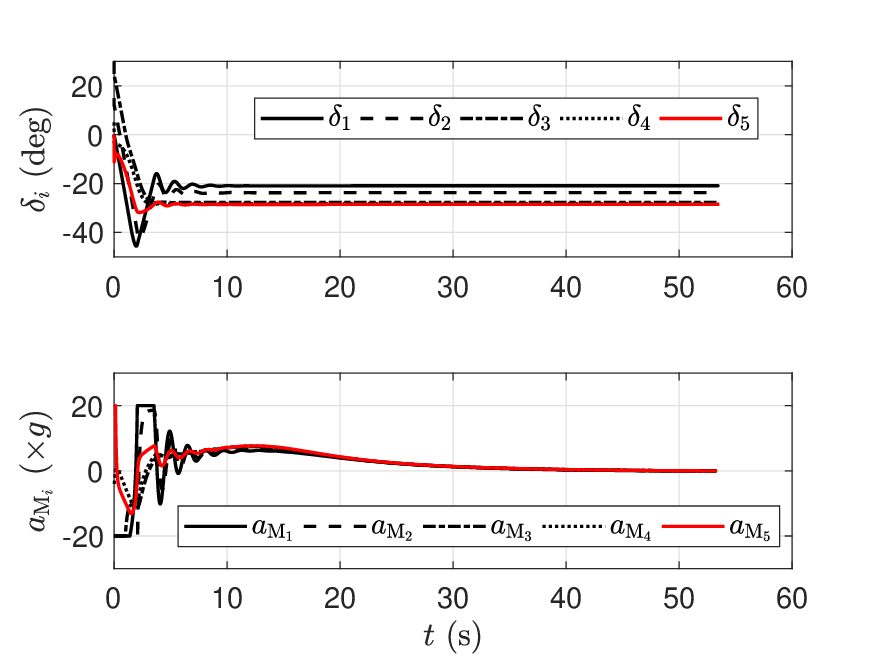}
		\caption{Deviation angles and lateral accelerations.}
	\end{subfigure}
	\begin{subfigure}[t]{.33\textwidth}
		\centering
		\includegraphics[width=\textwidth]{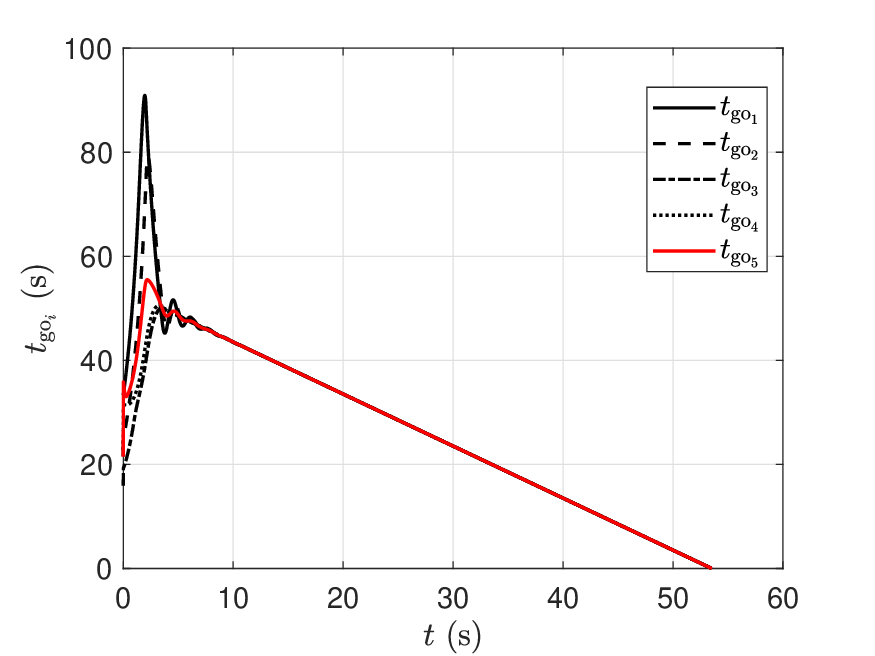}
		\caption{Time-to-go.}
	\end{subfigure}
	\caption{Impact time larger than the maximum initial time-to-go (pseudo-undirected cycle).}
	\label{fig:cycleHeNeg}
\end{figure*}
\begin{figure*}[h!]
	\centering
	\begin{subfigure}[t]{.33\textwidth}
		\centering
		\includegraphics[width=\textwidth]{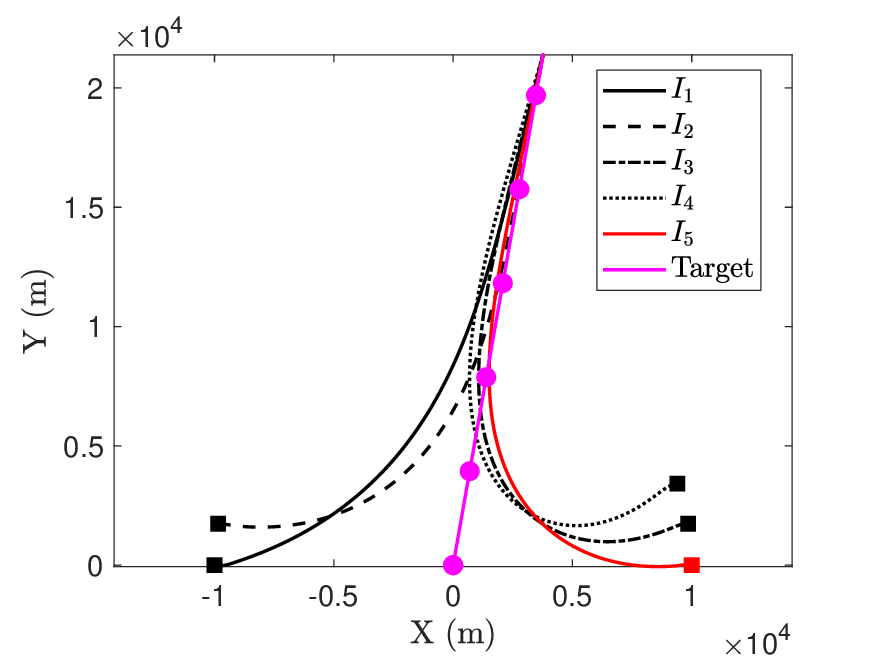}
		\caption{Trajectories.}
	\end{subfigure}
	\begin{subfigure}[t]{.32\textwidth}
		\centering
		\includegraphics[width=\textwidth]{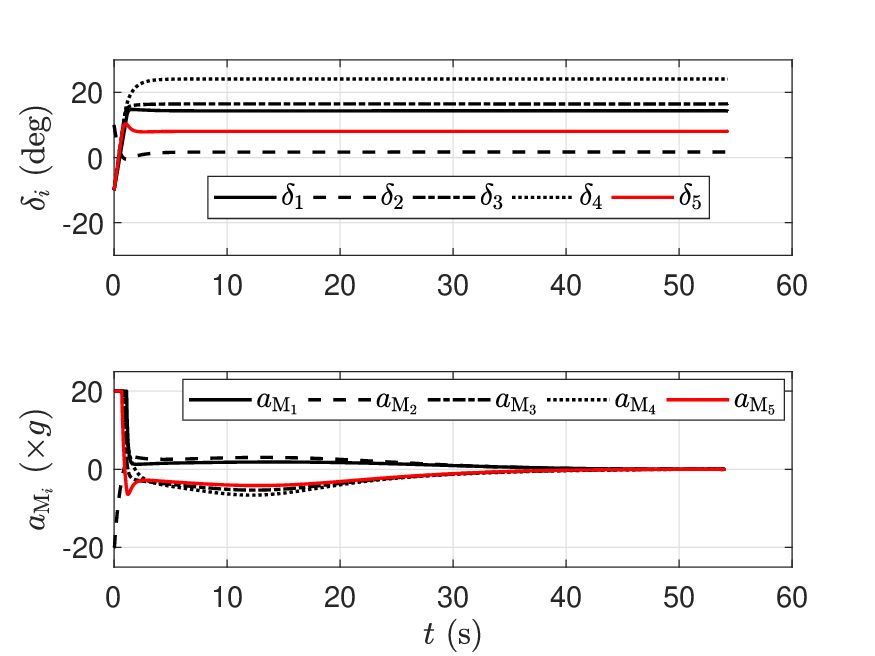}
		\caption{Deviation angles and lateral accelerations.}
	\end{subfigure}
	\begin{subfigure}[t]{.33\textwidth}
		\centering
		\includegraphics[width=\textwidth]{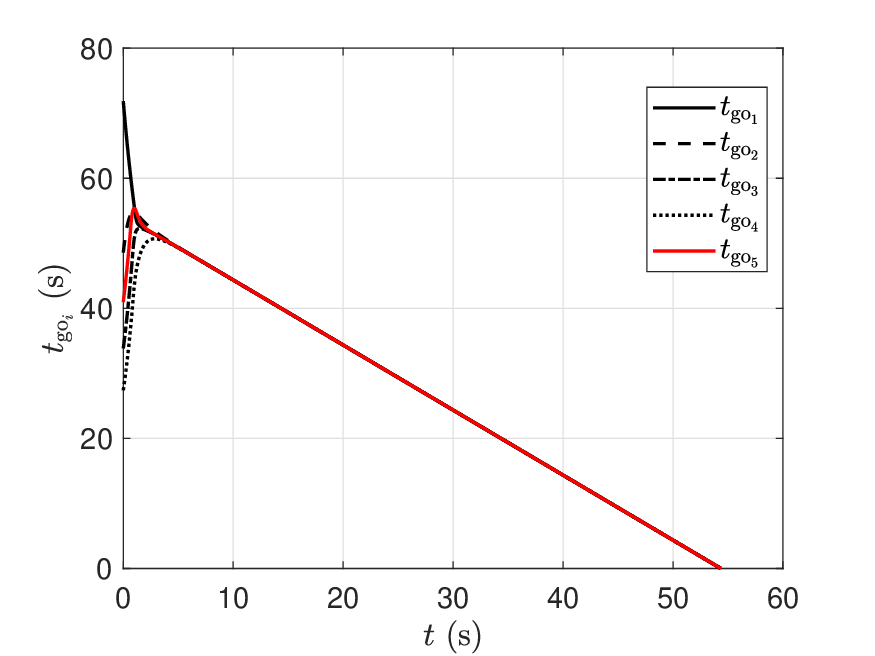}
		\caption{Time-to-go.}
	\end{subfigure}
	\caption{Weighted consensus in time-to-go (pseudo-undirected star).}
	\label{fig:starHePos}
\end{figure*}
\begin{figure*}[h!]
	\centering
	\begin{subfigure}[t]{.33\textwidth}
		\centering
		\includegraphics[width=\textwidth]{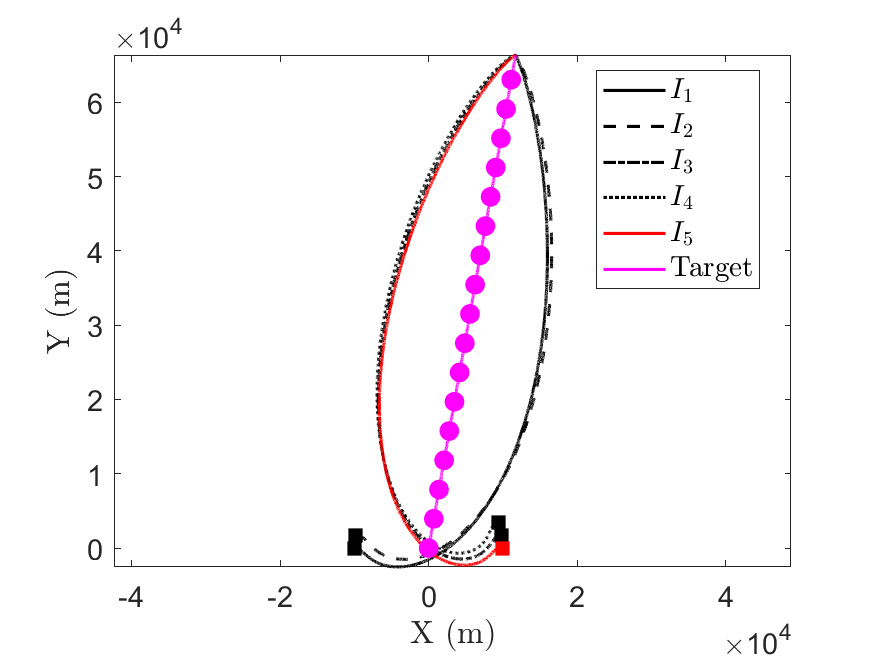}
		\caption{Trajectories.}
	\end{subfigure}
	\begin{subfigure}[t]{.32\textwidth}
		\centering
		\includegraphics[width=\textwidth]{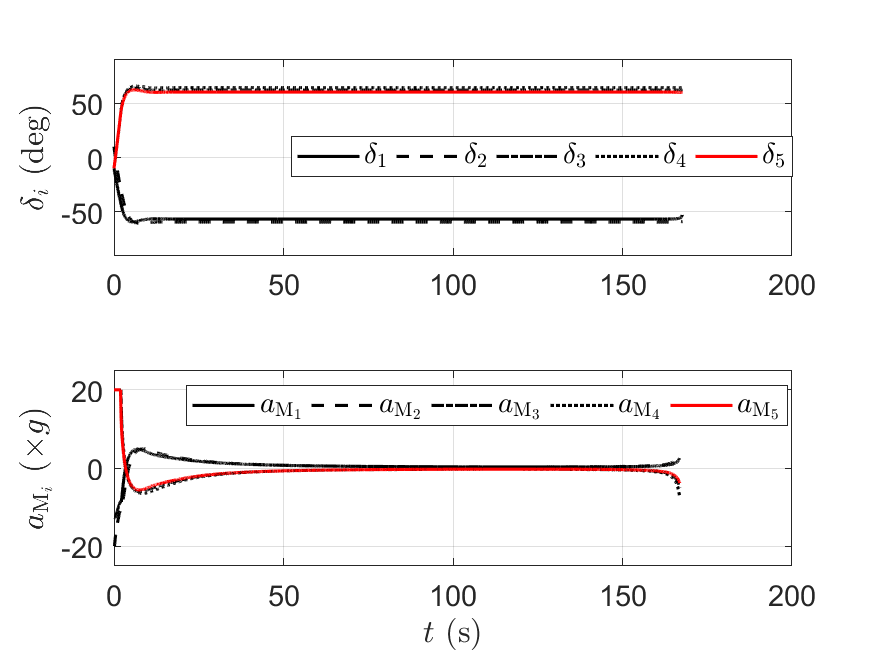}
		\caption{Deviation angles and lateral accelerations.}
	\end{subfigure}
	\begin{subfigure}[t]{.33\textwidth}
		\centering
		\includegraphics[width=\textwidth]{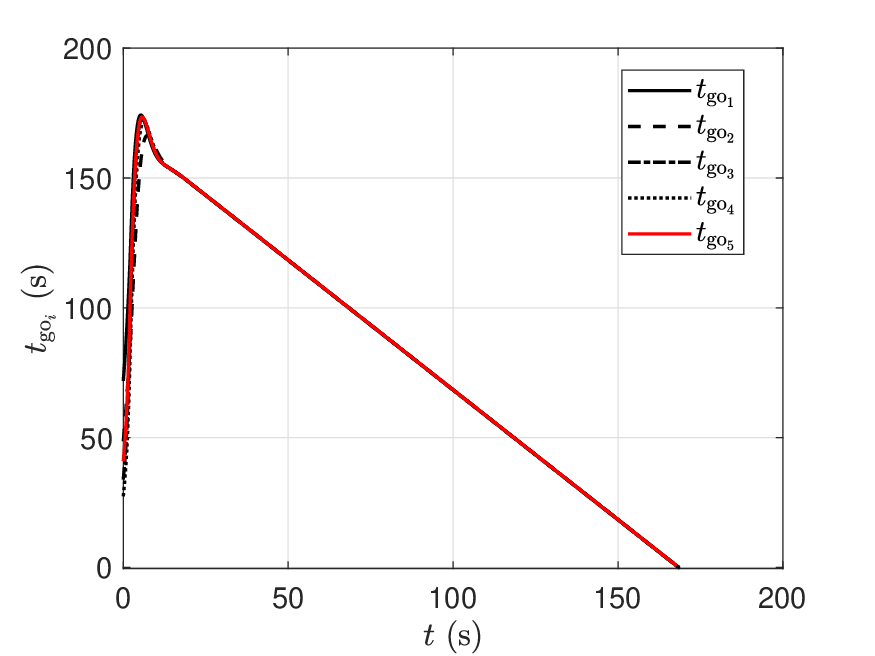}
		\caption{Time-to-go.}
	\end{subfigure}
	\caption{Impact time larger than the maximum initial time-to-go (pseudo-undirected star).}
	\label{fig:starHeNeg}
\end{figure*}

From the initial values of the interceptors' time-to-go given in \Cref{tb:initCycle}, the average time-to-go is computed as $22.92$ s, while the impact time is $30.74$ s. With heterogeneous positive weights, the impact time cannot be more than $33.33$ s and less than $15.92$ s. However, if the target is to be intercepted at a higher impact time, even higher than the largest initial time-to-go, then one of the edge weights can be perturbed to result an impact time (the consensus value) larger than $33.33$ s. With the edge weights given in \Cref{tb:conensusheterogeneousnominal}, we analyse the gain margin on the edge $e_{12}$ using the transfer function obtained by using \Cref{lem:TF}, which is given as 
\begin{equation}
	-M(s) = \dfrac{s^3 + 15.75s^2 + 47.44 s + 24.21}{s^4 + 32.75 s^3 + 323.7 s^2 + 1048 s + 1012}.
\end{equation}
This transfer function has multiple phase-crossover frequencies at $\omega_{pc}=0$ and $\omega_{pc}=-3.89$ rad/s with gain margins of $41.78$ and $17.08$, respectively. Hence, the effective gain margin is $17.08$, and the edge weight, $w_{12}$, can be perturbed up to $-10.08$ without missing the target. We perturb $w_{12}$ to $-8.51$, keeping rest of the edge weights in \Cref{tb:conensusheterogeneousnominal} unchanged. With this  change in an edge weight, the impact time (the consensus value) becomes $53.48$ s. This is illustrated in \Cref{fig:cycleHeNeg} with initial conditions same as that in \Cref{fig:cycleHePos}. Evidently, the proposed guidance strategy has enhanced applicability in simultaneous interception scenarios.

In a pseudo-undirected star graph (shown in \Cref{fig:PUGstar}), each interceptor is connected to a central interceptor. To intercept a target moving with a heading of $80^\circ$, interceptors communicate over such a topology, for which the edge weights and the initial engagement geometry are presented in \Cref{tb:wijStar,tb:initStar}, respectively. With these values, the average time-to-go is $44.53$ s, but the consensus in time-to-go is achieved for $54.31$ s, as portrayed in \Cref{fig:starHePos}. 
Using \eqref{eq:TF}, the transfer function corresponding to perturbation on $e_{15}$ is
\begin{equation}
	-M(s) = \dfrac{s^3 + 5.95 s^2 + 10.51 s + 4.788}{s^4 + 13.93 s^3 + 52.74 s^2 + 72.1 s + 31.73},
\end{equation}
with a gain margin of $6.62$. Making $w_{15}=-5.25$ results in a common interception at $168.38$ s, which is indeed quite larger than the maximum initial time-to-go ($71.84$ s). This is illustrated in \Cref{fig:starHeNeg}, confirming that the interval of achievable impact time values can be drastically enhanced.
\begin{table}[h!]
	\caption{Edge weights for pseudo-undirected star graph.}\label{tb:wijStar}
	\centering
	\begin{tabular}{|c|c|c|c|c|c|c|c|c|}
		\hline
		$w_{12}$  & $w_{13}$ &$w_{14}$ & $w_{15}$ & $w_{21}$  & $w_{31}$ &$w_{41}$ & $w_{51}$ \\ \hline 
		$0.7$ & $1.4$ & $1$ & $0.23$ & $2.4$ & $2.85$ & $4$ & $1.35$ \\  \hline
	\end{tabular}	
\end{table}
\begin{table}[ht!]
	\caption{Parameters of simulation for pseudo-undirected star graph.}\label{tb:initStar}
	\centering
	\begin{tabular}{|c|c|c|c|c|c|}
		\hline
		\phantom{-}& $I_1$  &$I_2$ & $I_3$ & $I_4$ & $I_5$\\ \hline 
		$\theta_i$ & $0^\circ$ & $-10^\circ$& $-170^\circ$& $-160^\circ$ & $180^\circ$\\  \hline
		$\gamma_{\mathrm{M}_i}$ & $-10^\circ$ & $0^\circ$ & $180^\circ$ & $190^\circ$ & $170^\circ$\\  \hline
		$t_{\mathrm{go}_i}(0)$ & $71.84$ s & $48.57$ s & $33.84$ s & $27.40$ s & $40.97$ s \\ \hline
	\end{tabular} 
\end{table}

Note that in each of these cases, we have tried to tailor the impact time (the consensus value) towards the higher initial time-to-go, or above the average value. Choosing an impact time as the average of the initial time-to-go, or even smaller, may have some serious repercussions over the guidance strategy. In practice, the lateral acceleration is usually limited. If a smaller time-to-go is desired, interceptors requiring higher time-to-go in the beginning may demand very high lateral acceleration values. This will eventually lead to actuator saturation, or even failure, which may disrupt the salvo mission. Thus, by a judicious choice of edge weights, the proposed cooperative guidance strategy offers the choice of a suitable impact time.

\section{Conclusions}\label{sec:conclusion}
In this paper, we proposed a cooperative guidance strategy to simultaneously intercept a moving target using deviated pursuit guidance.  We viewed the swarm communication topology  as a bi-directed network whose strength of interaction (edge weights) in forward and reverse directions were allowed to be different. Since we considered nonlinear engagement kinematics and could employ exact time-to-go, errors arising out of small angle assumption, and estimation of any kind were avoided. Prediction of any kind, such as predicting the target's interception point was not needed either due to the very nature of the guidance strategy, thereby reducing the computational complexity. We used heterogeneous positive edge weights in the communication network to obtain a weighted average consensus in interceptors' time-to-go values, while a slight negative perturbation in one of the edges helped in achieving significantly higher common impact time values for the group of cooperating interceptors. We also analysed the bound on the negative value of the chosen edge weight for some special topologies, and evaluated the consensus value as a function of the edge weights. This scheme helped in reducing the chances of interceptors' lateral accelerations reaching the saturation limits.  Interception of multiple targets, and dealing with unknown target manoeuvres are some of the potential extensions of this work. Consideration of aerodynamic parameter variation such as variable mass, thrust, drag and velocity, would be another interesting extension of this work.

\bibliographystyle{IEEEtran}
\bibliography{referencescopy}

\end{document}